\documentclass[margin,usenatbib]{mn2e}
\usepackage{amssymb,amsmath,graphicx,color}
                 
\newcommand{\Jvec}{\textit{\textbf{J}} }

\newcommand{\wvec}{\textit{\textbf{w}} }
\newcommand{\zvec}{\textit{\textbf{z}} }

\newcommand{\pvec}{\textit{\textbf{p}} }
\newcommand{\qvec}{\textit{\textbf{q}} }

\newcommand{\vvec}{\textit{\textbf{v}} }

\newcommand{\Fvec}{\textit{\textbf{F}} }
\newcommand{\gvec}{\textit{\textbf{g}} }

\newcommand{\lvec}{\textit{\textbf{l}} }
\newcommand{\Zvec}{\textit{\textbf{Z}} }
\newcommand{\Evec}{\textit{\textbf{E}} }

\newcommand{\Fmat}{\textsf{\textbf{F}} }

\newcommand{\Umat}{\textsf{\textbf{U}} }

\newcommand{\Amat}{\textsf{\textbf{A}} }

\newcommand{\Cmat}{\textsf{\textbf{C}} }
\newcommand{\Imat}{\textsf{\textbf{I}} }
\newcommand{\Jmat}{\textsf{\textbf{J}} }
\newcommand{\Mmat}{\textsf{\textbf{M}} }

\newcommand{\Wmat}{\textsf{\textbf{W}} }
\newcommand{\Vmat}{\textsf{\textbf{V}} }

\newcommand{\avec}{\textit{\textbf{a}} }
\newcommand{\bvec}{\textit{\textbf{b}} }
\newcommand{\cvec}{\textit{\textbf{c}} }
\newcommand{\dvec}{\textit{\textbf{d}} }
\newcommand{\Rvec}{\textit{\textbf{R}} }
\newcommand{\Gvec}{\textit{\textbf{G}} }

\newcommand{\dif}{{\rm d}}
\newcommand{\bnabla}{\mbox{\boldmath $\nabla$}}
\newcommand{\bfzeta}{\mbox{\boldmath $\zeta$}}
\newcommand{\mum}{\,\mu\mbox{m}}
\newcommand{\pc}{\,\mbox{pc}}

\begin{document}

\title[Density waves in near-Keplerian discs]{Density waves in debris discs
  and galactic nuclei}
\author[M.A. Jalali and S. Tremaine]
  {Mir Abbas~Jalali$^{1,2}$\thanks{mjalali@sharif.edu (MAJ)} and
   Scott Tremaine$^{2}$\thanks{tremaine@ias.edu (ST)} 
   \vspace{0.1cm} \\
   $^1$ Sharif University of Technology, Azadi Avenue, Tehran, Iran \\
   $^2$ School of Natural Sciences, Institute for Advanced Study, 
   Einstein Drive, Princeton, NJ 08540, U.S.A.}
   
\maketitle

\begin{abstract}
  We study the linear perturbations of collisionless near-Keplerian
  discs. Such systems are models for
  debris discs around stars and the stellar discs surrounding
  supermassive black holes at the centres of galaxies. Using a
  finite-element method, we solve the linearized collisionless
  Boltzmann equation and Poisson's equation for a wide range of
  disc masses and rms orbital eccentricities to obtain the
  eigenfrequencies and shapes of normal modes.  We find that these
  discs can support large-scale `slow' modes, in which the frequency is
  proportional to the disc mass. Slow modes are present for
  arbitrarily small disc mass so long as the self-gravity of the disc
  is the dominant source of apsidal precession. We find that slow
  modes are of two general types: parent modes and hybrid child modes,
  the latter arising from resonant interactions between parent
  modes and singular van Kampen modes. The most prominent slow modes
  have azimuthal wavenumbers $m=1$ and $m=2$.  We illustrate how slow
  modes in debris discs are excited during a fly-by of a
  neighbouring star. Many of the non-axisymmetric features seen in
  debris discs (clumps, eccentricity, spiral waves) that are commonly
  attributed to planets could instead arise from slow modes; the two hypotheses
  can be distinguished by long-term measurements of the pattern speed
  of the features. 
\end{abstract}



\section{Introduction}
Debris discs are planetesimal discs that are detected through
thermal infrared emission or scattered starlight from dust formed in 
recent planetesimal collisions. The bolometric luminosity from detectable 
debris discs is typically $\gtrsim 10^{-5}$ of the stellar luminosity, the 
inferred dust masses are typically $\lesssim 1 M_\oplus$, and the ages of 
the host stars range from 10 Myr to 10 Gyr (see \citealt{wy08} for a review). 

A variety of features in debris discs have been interpreted as evidence for planets. These include
structures in the $\beta$ Pictoris disc, including a warp
\citep{heap00}, a system of tilted rings \citep{wah03} and a bright
clump \citep{tel05}; clumps in the discs around Vega \citep{wy03},
$\epsilon$ Eridani \citep{gre05}, $\eta$ Corvi \citep{wy05},
and HD 107146 \citep{cor09}; the eccentricity of the 
discs around HR 4796A and Fomalhaut \citep{tel00,KGC05}; spiral
structure in the disc around HD 141569 \citep{cla03}; and sharp inner or
outer edges in the discs around Fomalhaut and HD 92945
\citep{KGC05,gol11}.

Detailed dynamical models have shown that most or all of these
features can be produced by planets (see \citealt{wy09} for a
review). Moreover, in the case of $\beta$ Pictoris \citep{lag10}, and
perhaps Fomalhaut \citep{K08}, planets have been detected that may
indeed be responsible for some or all of these features. Nevertheless,
it is important to ask what long-lived structures could arise in
debris discs {\em without} planets.

In this paper we examine the possibility that low-mass discs can
support long-lived normal modes maintained by the self-gravity of the
disc.  Normally it is assumed that debris discs cannot support such
modes because of their small masses; all localized disturbances are
dispersed by the Keplerian shear. However, a special feature of
Keplerian orbits is that eccentric orbits do not precess. Thus the
evolution of eccentric disturbances in a debris disc is governed by
the non-Keplerian forces, however small these may be. In this paper we
shall focus on the non-Keplerian forces arising from the self-gravity
of the disc.  We neglect other possible perturbations for a variety of
reasons. We ignore gravitational forces from planets because our
principal goal is to understand the properties of discs in the
simplest case, when no planets are present. We ignore radiation
pressure, even though this affects the dynamics of the dust that
dominates the thermal infrared emission and the scattered light; our
justification is that the large planetesimals that generate the dust
are unaffected by radiation pressure but we recognize that the
distribution of (invisible) parent bodies and (visible) dust is likely
to be different.  We ignore gas drag since old debris discs contain
little or no gas, and since the planetesimals are likely to be large
enough to be insensitive to drag.  We ignore collisions between
planetesimals because they are likely to be rare; indeed such
collisions probably drive the long-term erosion of the disc in which
case the collision time cannot be much less than the disc age.

Debris discs are distinct from protoplanetary discs: the latter
are comprised mostly of gas, not dust or planetesimals; they are much younger
(typically less than a few Myr) and more massive
(0.001--0.1$\,M_\odot$) than debris discs (see \citealt{wc11} for a
review). Protoplanetary discs are depleted by various processes,
including photoevaporation, accretion onto the host star, condensation
of refractory elements into dust grains and then planetesimals,
and stellar winds. Eventually they are likely to evolve into planetesimal/debris
discs. Although our analysis here is restricted to collisionless
systems, many of our results---in particular the existence of stable,
slow, lopsided modes supported by the self-gravity of the disc---also
apply to protoplanetary gas discs and may explain some non-axisymmetric
features of these discs. 

To summarize we treat debris discs as collisionless systems composed
of particles influenced only by the gravity of the host star and the
self-gravity of the disc. Their dynamics is therefore similar to the
dynamics of discs of stars orbiting near the supermassive black holes
found in the centres of most galaxies. Examples of these include the disc(s) of
young stars found in the central parsec of the Milky Way
\citep{genzel}, the two discs---one of young stars at $\sim0.1\pc$ and
one of old stars at $\sim1\pc$---found at the centre of M31
\citep{bk05}, and the stellar discs that are inferred to form in the
outer parts of quasar accretion discs \citep{good03}

The properties of the normal modes of  low-mass near-Keplerian discs were investigated
by Tremaine (2001; hereafter T01), who found that (i) the frequency of
the mode is proportional to the ratio $\mu$ of the masses of the disc
and central star, but the shape of the mode is independent of $\mu$ so
long as $\mu\ll1$ (hence these are called `slow' modes); (ii) all
slow modes are stable; (iii) in discs with rms eccentricity $e_{\rm
  rms}\ll1$ all slow modes have azimuthal wavenumber $m=1$, i.e., they
are lopsided.

The results in T01 are based on linear normal-mode
calculations for discs composed of particles in circular orbits, with
softened self-gravity used to mimic the effects of the velocity
dispersion or non-zero eccentricities of the particles. These
calculations are supplemented by analytic results using the WKB (short-wavelength)
approximation, which assumes that the wavelengths of the normal modes
are small compared to the radius. The WKB results appear to provide a
useful guide even though this short-wavelength approximation is not
realistic for some of the disc modes. In this paper the effects of the
velocity dispersion are computed directly, and we examine discs with a
range of rms eccentricities $e_{\rm rms}$, from nearly zero (`cold'
discs) to $\sim 0.35$ (`warm' discs). Our numerical results are
derived using a finite-element method (FEM) for studying the linear
normal modes of collisionless self-gravitating discs, as described in
\cite{J10}. In particular, we intend to
address the following questions: (i) What are the properties of the
frequency spectra of near-Keplerian discs? (ii) Are there any
unstable modes? (iii) Are there isolated oscillatory modes in the spectrum
that survive Landau damping? (iii) What are the differences between
the spectra of cold and warm discs? (iv) How can stable density waves
be excited in such discs?

We introduce a family of axisymmetric near-Keplerian discs in
\S\ref{sec:disc-model} and construct their equilibrium phase-space
distribution functions (DFs) in \S\ref{sec:DF}.  We obtain the
governing equations of the perturbed dynamics in \S\ref{sec:dynamics}
and explain the numerical solution procedure in
\S\ref{sec:FEM-model}. We present the frequency spectra of our discs
in \S\ref{sec:prograde-waves} and discuss the characteristics of
eigenmodes in warm and cold discs. We describe how these waves can be
excited by tidal forces in \S\ref{sec:excite}. The reader who is
mainly interested in the application of our results to debris discs
and galactic nuclei can focus on Figures \ref{fig7} and \ref{fig10}
and the discussion in \S\ref{sec:appl}.

\section{The Model}
\label{sec:disc-model}

\begin{figure}
\centerline{\hbox{\includegraphics[width=0.45\textwidth]{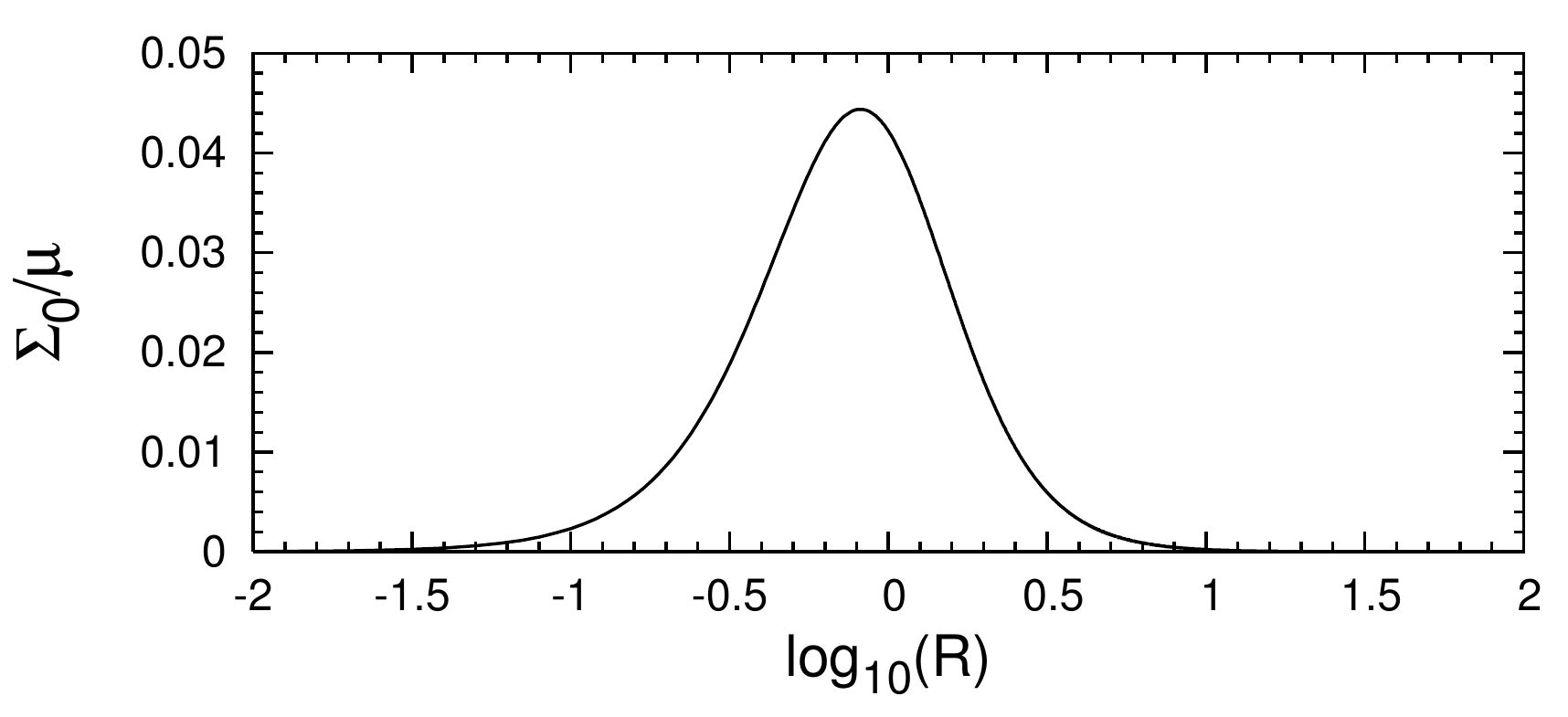} } }
\centerline{\hbox{\includegraphics[width=0.45\textwidth]{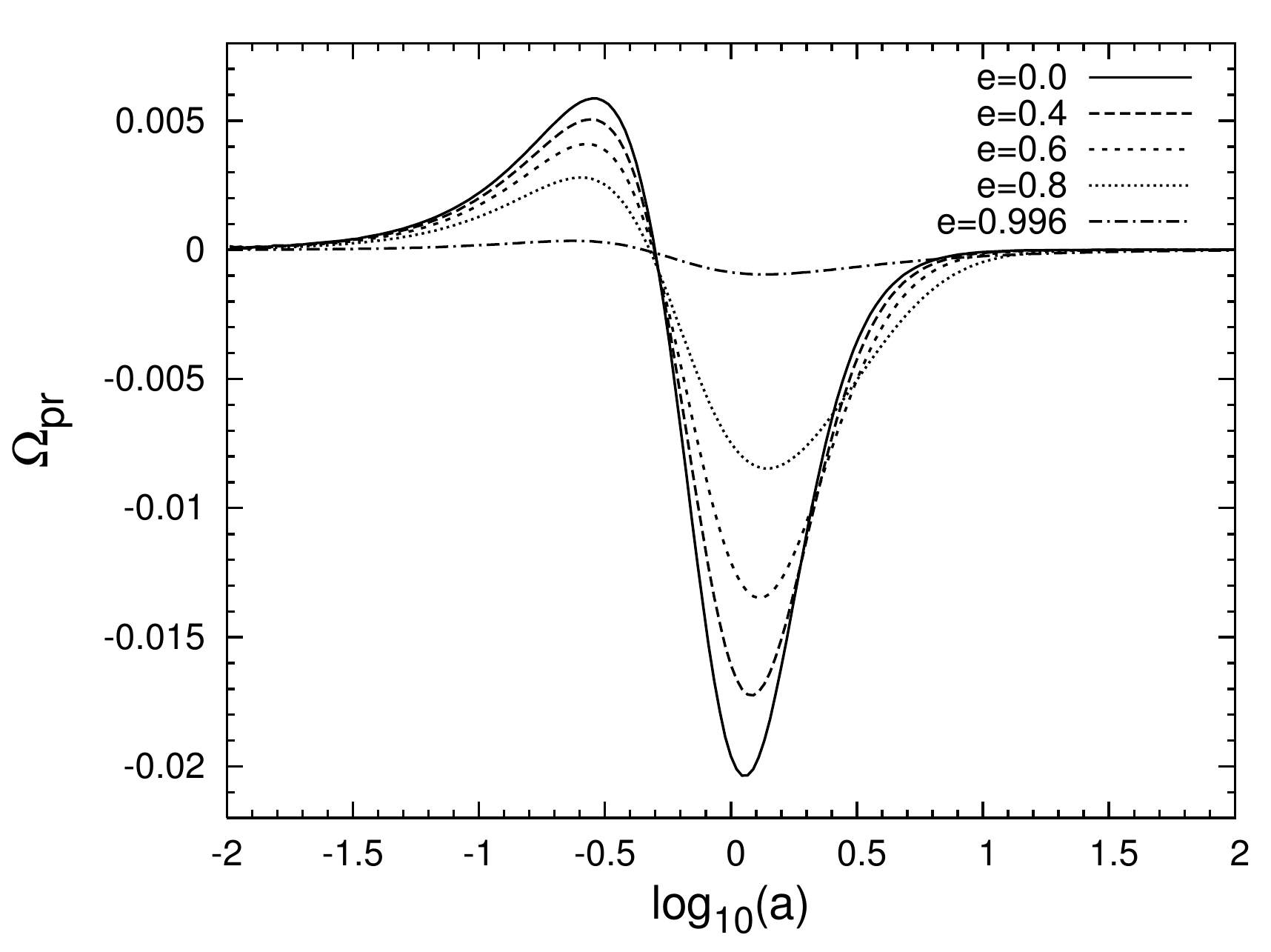} } }
\caption{{\it Top}: Surface-density profile of the composite Toomre
disc (eq.\ \ref{eq:sd}). {\it Bottom}: Variation of the precession rate 
$\Omega_{\rm pr}$ for $\mu=0.1$. For radial orbits (eccentricity $e=1$)
$\Omega_{\rm pr}=0$.}
\label{fig1}
\end{figure}

We introduce a simple model of annular discs 
around massive objects by subtracting two \cite{T63} discs with $n=1$ 
and $n=2$; the resulting surface density is 
\begin{align}
S_{\rm d}(r) &= \frac{3M_{\rm d}}{4\pi b^2}
\left \{ \frac{1}{[1+(r/b)^2]^{3/2}} - \frac{1}{[1+(r/b)^2]^{5/2}} \right \}, 
\notag \\
 &= \frac{3M_{\rm d}}{4\pi b^2} \frac{(r/b)^2}{[1+(r/b)^2]^{5/2}},
\label{eq:sd}
\end{align}
where $M_{\rm d}$ is the disc mass, $b$ is a length scale, and $r$ 
is the radial distance to the central star. The potential 
corresponding to the surface density 
$S_{\rm d}$ is 
\begin{equation}
\Phi_{\rm d}(r)=-\frac{GM_{\rm d}}{2b} \frac{1+2(r/b)^2}{[1+(r/b)^2]^{3/2}},
\end{equation}
with $G$ being the gravitation constant. For a central 
star of mass $M_{\star}$, the total potential governing the motion 
of particles is 
\begin{eqnarray}
\Phi_0(r)=-\frac{GM_{\star}}{r}+\Phi_{\rm d}(r).
\end{eqnarray}
We define  
\begin{eqnarray}
\mu = \frac{M_{\rm d}}{M_{\star}},~~R=r/b,
\end{eqnarray}
and work with the dimensionless unperturbed potential
\begin{eqnarray}
V_0(R)\equiv\frac{b \Phi_0}{G M_{\star}}=-\frac{1}{R} - \frac{\mu}{2} \frac{1+2 R^2}{(1+R^2)^{3/2}},
 \label{eq:V0}
\end{eqnarray}
and density 
\begin{eqnarray}
\Sigma_0(R) \equiv \frac{b^2 S_{\rm d}}{M_{\star}}= 
\frac{3 \mu}{4\pi} \frac{R^2}{(1+R^2)^{5/2}}.
 \label{eq:Sigma0}
\end{eqnarray}
The top panel in Figure \ref{fig1} shows the radial profile of $\Sigma_0/\mu$.

The velocity of circular orbits, $v_c(R)$, is determined from 
\begin{eqnarray}
v_c^2(R)=R\frac{\dif V_0}{\dif R} = \frac 1R +\frac{\mu}{2} 
\frac{R^2(2R^2-1)}{(1+R^2)^{5/2}}.
\label{eq:circular-orbit-v}
\end{eqnarray}
The second term on the right side of (\ref{eq:circular-orbit-v}) becomes negative for $R^2<1/2$. This
means that our discs cannot exist in the absence of a central point mass.
More precisely, $v_c^2\ge0$ at all $R$ if and only if $\mu \le 5^{5/2}$; 
this is not a limitation in practice since protoplanetary discs are expected to 
have $\mu\ll1$. 

We restrict ourselves to razor-thin discs since the vertical structure of thin
discs should not strongly affect their large-scale response. Using the polar
coordinates $(R,\phi)$ and their corresponding generalized momenta
$(p_R,p_{\phi})$, the Hamiltonian function governing the motion of particles
reads
\begin{eqnarray}
{\cal H}_0 (p_R,p_\phi,R)\equiv E = \frac{p_R^2}{2}+\frac{p_{\phi}^2}{2R^2}+V_0(R).
\label{eq:hhhh}
\end{eqnarray}
Since $\phi$ is a cyclic coordinate, its conjugate momentum $p_{\phi}$ is a
constant of motion in the unperturbed disc. The orbital energy $E$ is another
integral of motion. Canonical perturbation theories describing the motion of
particles, and the perturbed collisionless Boltzmann equation, are
substantially simplified by using the action
variables $\Jvec=(J_R,J_{\phi})$ and their conjugate angles $\wvec
=(w_R,w_{\phi})$ with
\begin{eqnarray}
J_R=\frac{1}{2\pi}\oint p_R \,\dif R,~~
J_{\phi}=\frac{1}{2\pi} \oint p_{\phi} \,\dif \phi=p_{\phi}.
\label{eq:define-actions}
\end{eqnarray}
These integrals are taken along the orbits, which consist of slowly precessing
Kepler ellipses when $\mu\ll1$.  The unperturbed Hamiltonian ${\cal H}_0$ depends
only on the actions, not the angles. The action $J_{\phi}=L$ 
is the magnitude of the angular-momentum vector. In the angle-action space, 
the equations of motion become  
\begin{eqnarray}
\dot \wvec ={\bf \Omega}(\Jvec)=\frac{\partial {\cal H}_0(\Jvec)}{\partial \Jvec},
~~ \dot{\!\!\Jvec} =0,
\end{eqnarray}
and the orbital frequencies ${\bf \Omega}(\Jvec)=(\Omega_R,\Omega_{\phi})$ 
are computed from 
\begin{eqnarray}
\frac{2\pi}{\Omega_R(\Jvec)} = \oint \frac{\dif R}{ p_R(R,\Jvec) } ,~~
\frac{\Omega_{\phi}(\Jvec)}{\Omega_R(\Jvec)} =
\frac{J_{\phi}}{2\pi}\oint \frac{\dif R}{ R^2 p_R(R,\Jvec) } .
\label{eq:orbfreq}
\end{eqnarray}
In the limit $\mu\rightarrow 0$, the potential is Keplerian and we have
$\Omega_R=\Omega_{\phi}=a^{-3/2}$ with $a$ being the orbital semi-major
axis. For $0<\mu\ll1$ the radial and azimuthal frequencies are no longer equal,
but their difference $\Omega_{\rm pr}=\Omega_{\phi}-\Omega_R$ is small, and
that is the precession rate of the line of apsides. The Taylor expansion of
$\Omega_{\rm pr}$ begins with terms of ${\cal O}(\mu)$ because $\Omega_{\rm
  pr}$ vanishes for Keplerian orbits. Consequently, for $\mu\ll 1$, the
precession rate is proportional to the disc mass. For nearly circular
orbits, the precession rate is given analytically by
\begin{equation}
\Omega_{\rm
  pr}=\frac{3\mu}{4}\frac{R^{3/2}(1-4R^2)}{(1+R^2)^{7/2}}+{\cal O}(\mu^2).
\label{eq:precrate}
\end{equation}

Instead of the actions one may use the semi-major axis
$a(\Jvec)$ and eccentricity $e(\Jvec)$ defined by
\begin{eqnarray}
a =\frac{R_{\rm min}(\Jvec)+R_{\rm max}(\Jvec)}{2},~~
e =\frac{ R_{\rm max}(\Jvec)-R_{\rm min}(\Jvec) }
        { R_{\rm min}(\Jvec)+R_{\rm max}(\Jvec) },				
\end{eqnarray}
where $R_{\rm min}(\Jvec)$ and $R_{\rm max}(\Jvec)$ are the minimum and
maximum distances of particles from the central star. These definitions are
consistent with the standard Keplerian definitions when the disc mass
vanishes. In the bottom panel of Figure \ref{fig1}, we have plotted the
variation of $\Omega_{\rm pr}$ versus $a$ for $\mu=0.1$ and several choices
of $e$. It is seen that the precession rate of orbits---of any
eccentricity---has a positive peak within the region where $\Sigma_0$ is
rising, and then switches sign and remains negative in the outer regions. 
The precession rate crosses through zero near $a=0.5$ at all eccentricities. 
The maximum precession rate for nearly circular orbits and $\mu\ll1$ is given 
by equation (\ref{eq:precrate}) as $\omega_0=0.05861\mu$, which occurs at 
$R=0.2859$. In \S\ref{sec:prograde-waves}, we shall show that the pattern 
speeds of stable waves are closely related to $\omega_0$.

\section{Phase-space distribution function}
\label{sec:DF}

Particle orbits in collisionless discs are not necessarily circular. We therefore 
construct phase-space distribution functions (DFs) that enable us to 
distribute non-circular orbits in the disc. We seek DFs of the form
\citep{Sa88,PLB96}
\begin{eqnarray}
f_0({\cal E},L)=L^{2K+2} g_K({\cal E}),~~
{\cal E}=-E,
\label{eq:equilibrium-DF}
\end{eqnarray}
where $0\le L \le L_c({\cal E})$, $L_c({\cal E})$ is the angular
momentum of a circular orbit with energy $E=-{\cal E}$, and $K$ is a
positive integer.
To reproduce the surface density the DF must satisfy the relation 
\begin{eqnarray}
\Sigma_0(R) = 2 \int_{0}^{\Psi} \!\!\dif {\cal E} 
\int_{0}^{L_{\rm max}} \frac{f_0({\cal E},L)~\dif L}{\sqrt{L_{\rm max}^2-L^2}},
~~ \Psi=-V_0,
\label{eq:fundamental-equation}
\end{eqnarray}
where $L_{\rm max}=R[2(\Psi-{\cal E})]^{1/2}$. Substituting (\ref{eq:equilibrium-DF}) 
into (\ref{eq:fundamental-equation}) and performing the integral over $L$ we
find 
\begin{eqnarray}
2^{K+1}B\left (K \! + \! \frac 32,\frac 12 \right ) \int_{0}^{\Psi} \!\! \dif {\cal E} \, 
g_K({\cal E}) (\Psi-{\cal E})^{K+1} = \frac{\Sigma_0(R)}
{R^{2K+2}},
\label{eq:determining-eq-for-g}
\end{eqnarray}
where $B(p,q)$ is the beta function. Taking the $(K+2)$th-order derivative 
of both sides of (\ref{eq:determining-eq-for-g}) with respect to $\Psi$ gives 
an explicit analytic form of $g_K$,
\begin{eqnarray}
g_K(\Psi)=\frac{1}{\sqrt{\pi}2^{K+1} \Gamma(K+3/2)}
\frac{\dif ^{K+2}}{\dif \Psi^{K+2}}
\frac{\Sigma_0(R)} {R^{2K+2}} .
\end{eqnarray}

One needs to know explicitly the function $R(\Psi)$ before doing the derivatives on the right 
side of (\ref{eq:determining-eq-for-g}). Since $\mu$ is small in 
the discs we are considering, we utilize a perturbation method to compute $R$ in terms of $\Psi$. 
Let us define $u=1/R$ and rewrite (\ref{eq:V0}) in the form
\begin{eqnarray}
\Psi = u + \mu Q(u),~~ Q(u)=\frac{1}{2} \frac{u(2+u^2)}{(1+u^2)^{3/2}}.
\label{eq:Psi-vs-u}
\end{eqnarray}
We now assume a formal series expansion for $u$ in terms of $\mu$ as
\citep{B64}
\begin{eqnarray}
u(\Psi) = u_0(\Psi) + \mu u_1(\Psi) + \mu^2 u_2(\Psi) + \cdots,
\label{eq:expand-u}
\end{eqnarray}
and substitute this into (\ref{eq:Psi-vs-u}). The functions $u_j(\Psi)$ are recursively determined 
by putting equal to zero the coefficients of $\mu^j$ ($j=0,1,2,\cdots$). The recursion 
begins with $u_0=\Psi$. Up to the third-order terms, we obtain
\begin{align}
u_1 =& -Q(u_0),  \notag \\
u_2 =& -u_1 Q'(u_0), \notag \\
u_3 =& -u_2 Q'(u_0)-\frac {1}{2} u_1^2 Q''(u_0),
\end{align}
where $Q'(u)=\dif Q/\dif u$. The series for $u$ converges rapidly so keeping the 
terms of ${\cal O}(\mu^2)$ is quite sufficient for computing $R(\Psi)=1/u(\Psi)$ in discs 
with $\mu \le 0.1$.

\begin{figure}
\centerline{\hbox{\includegraphics[width=0.45\textwidth]{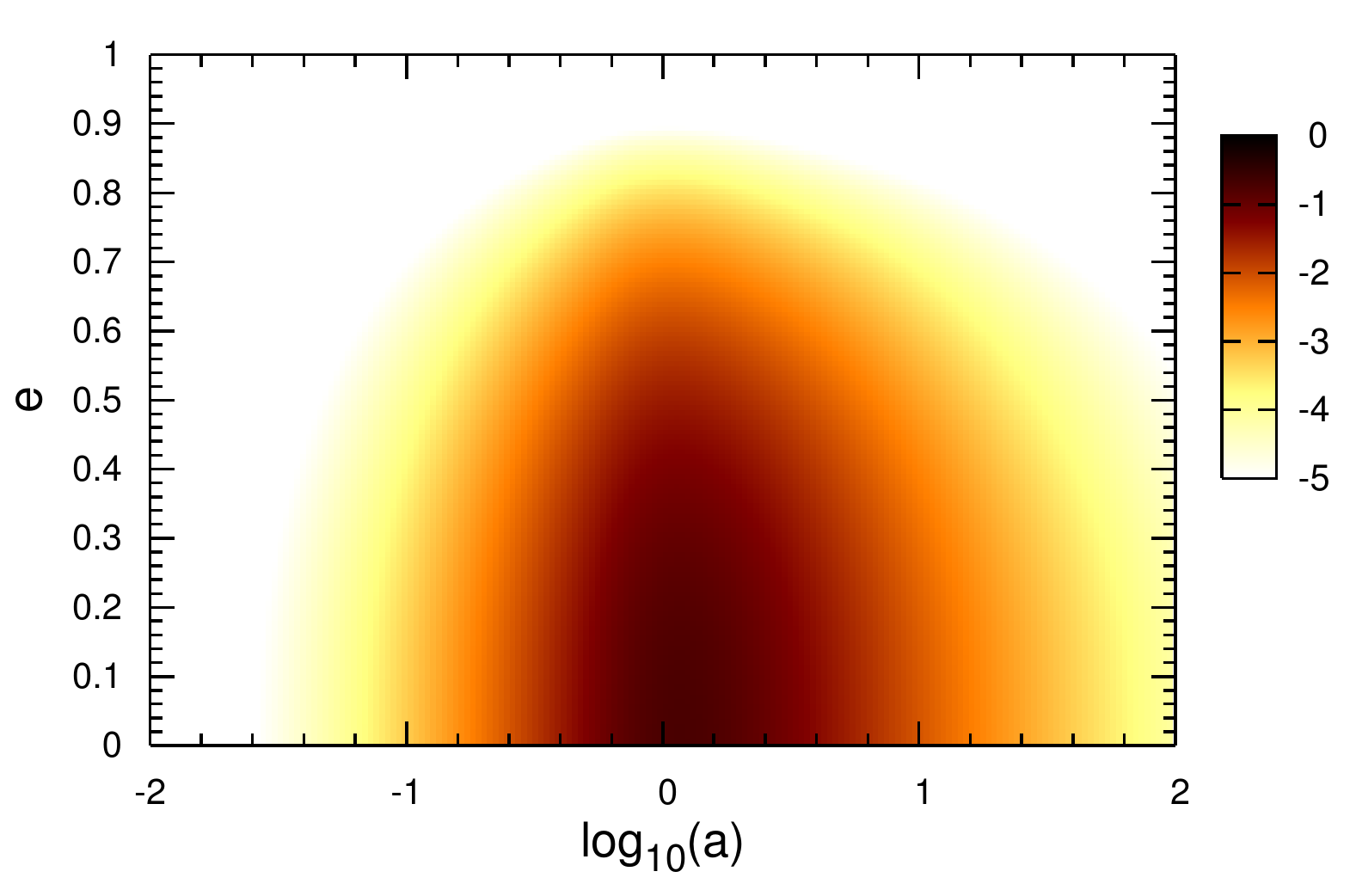}} }
\centerline{\hbox{\includegraphics[width=0.45\textwidth]{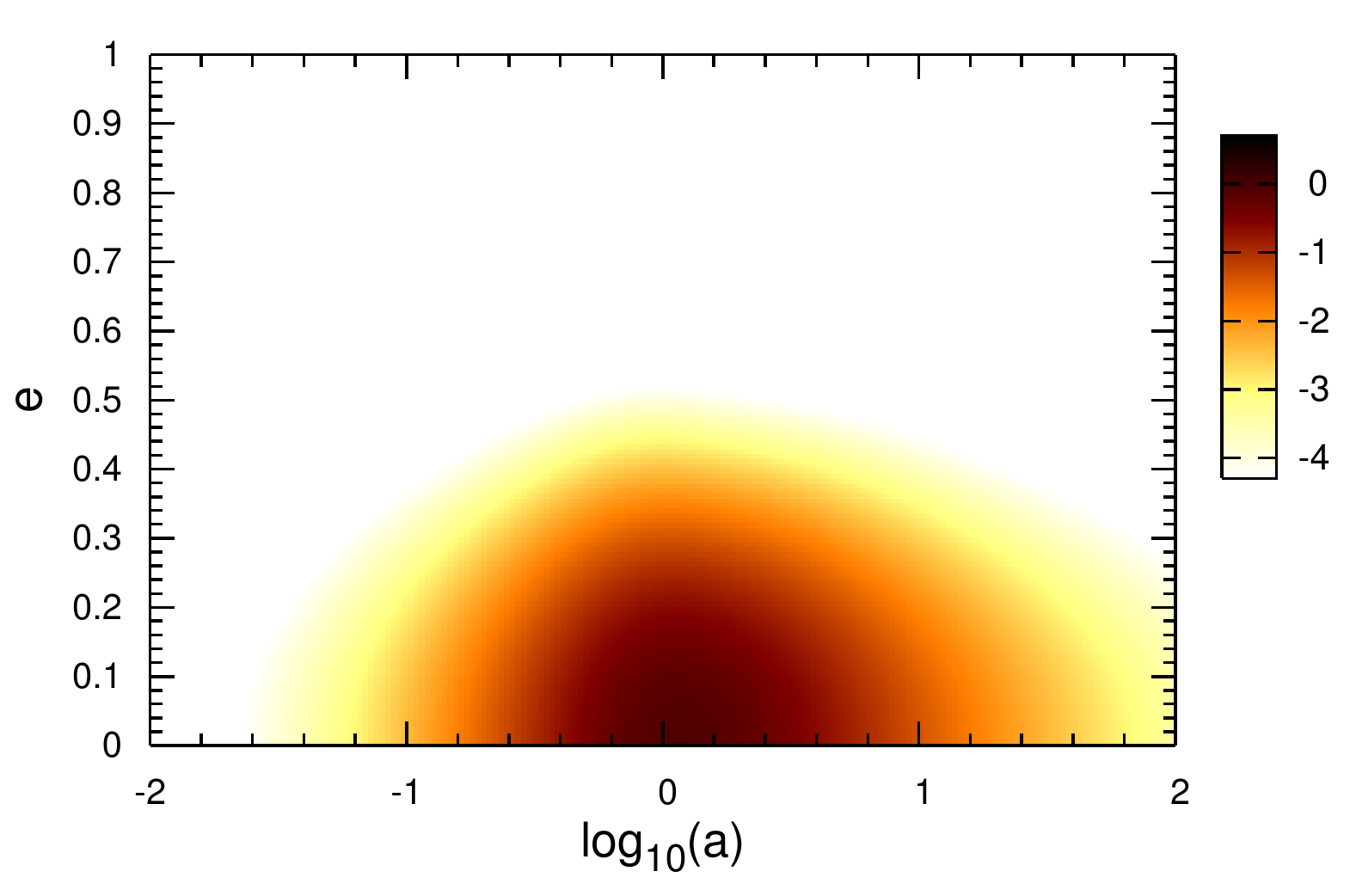}} }
\caption{Contours of $\log_{10}[f_0(a,e)/ \mu]$ for $\mu=0.1$. The maximum surface 
density $\Sigma_{0}(R)$ occurs at $R=0.8165$. Therefore, the highest phase-space 
density appears in the vicinity of $\log_{10}(a)\simeq  0$. {\it Top}: $K=5$. 
{\it Bottom}: $K=29$.}
\label{fig2}
\end{figure}

The functions $g_K({\cal E})$ admit negative values for $K=0,1$, but they are positive-definite 
and therefore physical for plausible values of $\mu <1$ when $K \ge 2$. We have plotted 
the contours of $\log_{10}(f_0/\mu)$ using $(a,e)$ as independent variables in Figure \ref{fig2} 
for $K=5$ and $K=29$. The mean and rms eccentricity of the disc, $\bar e$ and $e_\mathrm{rms}$, 
are given by 
\begin{align}
\bar e=&\frac{\int e f_0(\Jvec) ~ \dif^2 \Jvec}{\int  f_0(\Jvec) ~
  \dif^2
  \Jvec}=\frac{\Gamma(\frac{3}{2})\Gamma(\frac{5}{2}+K)}{\Gamma(3+K)}+{\cal O}(\mu)
\notag \\
e_\mathrm{rms}^2=&\frac{\int e^2 f_0(\Jvec) ~ \dif^2 \Jvec}{\int  f_0(\Jvec) ~
  \dif^2  \Jvec}=\frac{2}{2K+5}+{\cal O}(\mu). 
\label{eq:erms}
\end{align}
Larger values of $K$ correspond to colder discs.  For $\mu\ll1$ the mean eccentricity 
$\bar e=0.329$ for $K=5$ and $\bar e=0.159$ for $K=29$. When $K\gg1$ the DF at a 
given energy or semi-major axis approaches the Schwarzschild or Rayleigh DF, 
\begin{equation}
f_0(e^2)de^2\propto \exp(-e^2/e_0^2)de^2,  \quad e_0^{-2}=K+1/2,
\end{equation} 
In this limit the mean and rms eccentricity are related to $e_0$ by
$\bar e=\sqrt{\pi} e_0 / 2$, $e_{\rm rms}=e_0$. 

\begin{figure}
\centerline{\hbox{\includegraphics[width=0.45\textwidth]{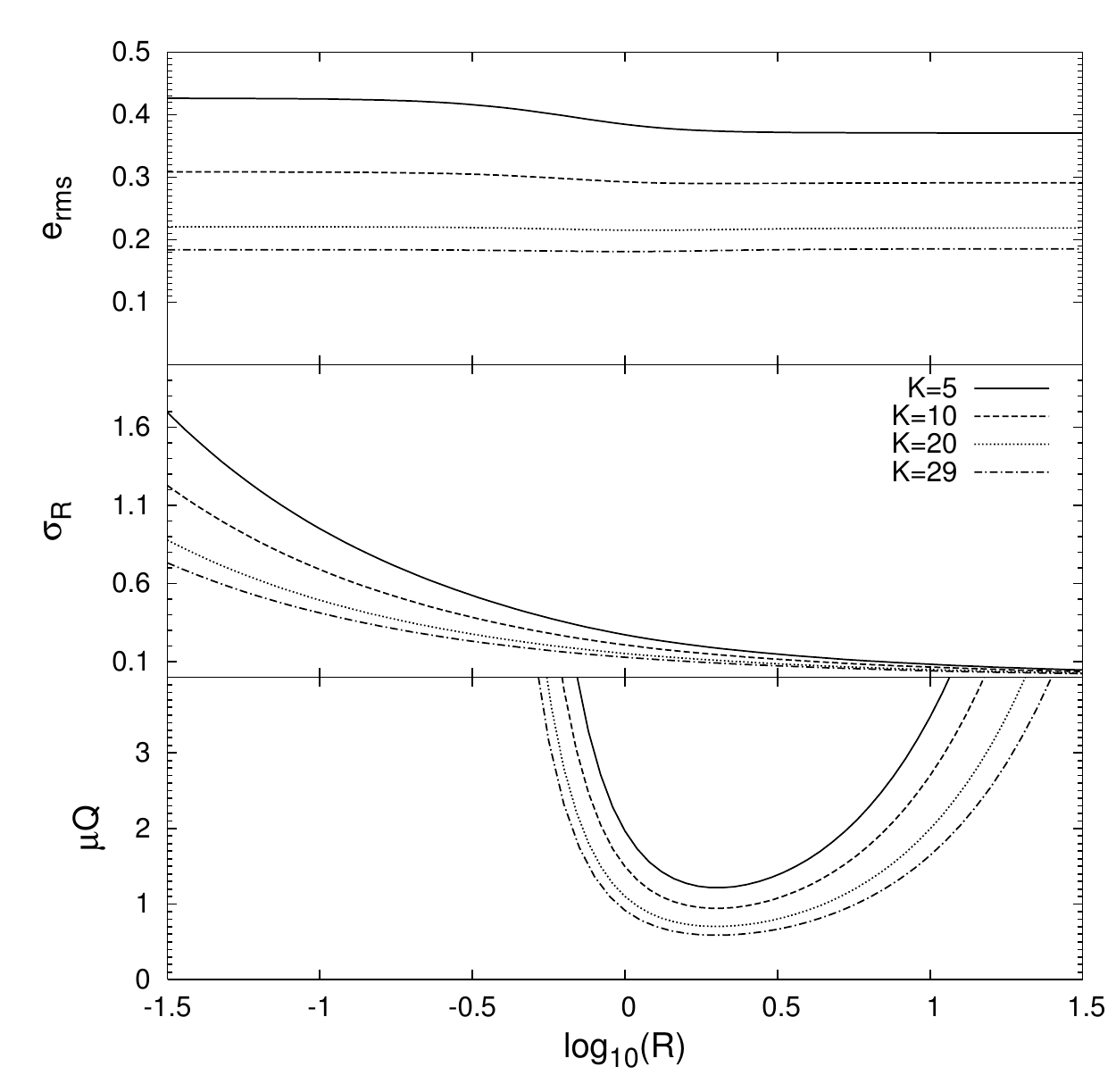}} }

\caption{The rms eccentricity, radial velocity dispersion, and $\mu Q$
  (the disc/star mass ratio times Toomre's stability parameter $Q$) as
  functions of radius. When $\mu\ll1$ all three plots are independent
  of $\mu$; the curves are from numerical models with $\mu=0.1$.}
\label{fig:esq}
\end{figure}

A necessary condition for stability to small-scale axisymmetric
disturbances is that Toomre's $Q>1$; here
$Q=\sigma_R\Omega_R/(3.36\Sigma_0)$ where $\sigma_R$ is the radial
velocity dispersion. The models in this paper with $\mu\ll1$ have
$Q>0.5/\mu$ everywhere and thus are stable in this sense.  The top two
panels of Figure \ref{fig:esq} show the rms eccentricity and
$\sigma_R$ as functions of radius; for $\mu\ll1$ these are independent
of $\mu$. The bottom panel shows $\mu Q$ which is also independent of
$\mu$ for $\mu\ll1$. Note in particular
that the rms eccentricity is almost independent of radius.

\section{Perturbed Dynamics} 
\label{sec:dynamics}
We assume that gas drag, collisions, and other non-gravitational
effects are negligible so the disc can be treated as a collisionless
fluid. We impose small-amplitude disturbances to the surface density,
potential, and DF:
\begin{align}
\Sigma(R,\phi,t) =& \Sigma_0(R) + 
\epsilon \Sigma_1(R,\phi,t), \label{eq:perturbed-sigma} \\
V(R,\phi,t) =&  V_0(R) + 
\epsilon V_1(R,\phi,t) + \epsilon V_{\rm e}(R,\phi,t), \label{eq:perturbed-pot} \\
f(\wvec,\Jvec,t) =&  f_0(\Jvec) +
\epsilon f_1(\wvec,\Jvec,t), \label{eq:perturbed-DF}
\end{align}
where $\epsilon\ll 1$ and $V_{\rm e}$ is an external perturbing
potential, perhaps induced by a binary companion, an encounter with a
passing star, or the tidal field of the birth cluster. The perturbed
surface density $\Sigma_1$ and its corresponding potential $V_1$ are
related through Poisson's integral: 
\begin{eqnarray}
V_1(R,\phi,t) \!\!\! &=& \!\!\! -G\iint \frac{\Sigma_1(R',\phi',t) 
R'\,\dif R'\,\dif \phi'}{\sqrt{R^2+R'^2-2RR'\cos(\phi-\phi')}} \nonumber \\
\!\!\! &+& \!\!\! G R \iint \frac{\Sigma_1(R',\phi',t) \cos(\phi-\phi') 
\,\dif R' \,\dif \phi'}{R'},
\label{eq:poisson-integral}
\end{eqnarray}
and we consider self-consistent density perturbations so that 
\begin{eqnarray}
\Sigma_1 = \int f_1 ~\dif^2 \vvec.
\label{eq:density-DF-relation}
\end{eqnarray}
The second term on the right side of (\ref{eq:poisson-integral}) is the indirect 
potential perturbation that arises because we are working in a
non-inertial reference frame centred on the star. It is non-zero only
for $m=1$ perturbations since perturbations with $m\not=1$ leave the
centre of mass of the disc unchanged.  
For a particle with actions $\Jvec$, the radial distance $R$ and $\exp({\rm i}m\phi)$ can 
be expanded as Fourier series in the angle variables, 
\begin{eqnarray}
R = \sum_{l=-\infty}^{+\infty} \xi_{l}(\Jvec) e^{{\rm i} l w_R}, 
~~
e^{{\rm i}m\phi} = e^{{\rm i}m w_{\phi} } \!\! \sum_{l=-\infty}^{+\infty} 
\!\! \eta_{l}(\Jvec) e^{{\rm i} l w_R}.
\label{eq:R-and-phi-vs-angle-actions}
\end{eqnarray}
Any function of $R$ and $\phi$ that is $2\pi$-periodic in $\phi$ can thus be
expressed in the $(\wvec,\Jvec)$ coordinates. For the Hamiltonian function
that governs the motion of particles, we write
\begin{eqnarray}
{\cal H} = {\cal H}_0(\Jvec)+\epsilon V_1(\wvec,\Jvec,t)+
\epsilon V_{\rm e}(\wvec,\Jvec,t).
\end{eqnarray}
where ${\cal H}_0$ is defined in equation (\ref{eq:hhhh}). 
Therefore, the perturbed equations of motion become
\begin{align}
\dot \wvec =&  \frac{\partial {\cal H}}{\partial \Jvec} 
= {\bf \Omega}(\Jvec)+\epsilon \frac{\partial}{\partial \Jvec}
\left ( V_1+V_{\rm e} \right ), \\
\dot {\!\!\Jvec} =& -\frac{\partial {\cal H}}{\partial \wvec}=
-\epsilon \frac{\partial}{\partial \wvec}
\left ( V_1+V_{\rm e} \right ).
\end{align}
It is obvious that the actions vary slowly in the perturbed disc. 
Subtracting the evolutionary equations of $w_{\phi}$ and $w_R$ gives
the apsidal precession rate in the perturbed disc, 
\begin{eqnarray}
\dot w_{\phi}- \dot w_R = \Omega_{\rm pr}(\Jvec)+\epsilon 
\left ( \frac{\partial}{\partial J_{\phi}} -\frac{\partial}{\partial J_R} 
\right )
\left ( V_1+V_{\rm e} \right ).
\label{eq:evolution-wphi-wR}
\end{eqnarray}
Since $\Omega_{\rm pr}={\cal O}(\mu)$, for low-mass discs ($\mu\ll1$) this
equation contains two small parameters, $\mu$ and $\epsilon$.

The DF in the perturbed disc obeys the collisionless Boltzmann
equation (CBE),
\begin{eqnarray}
\frac{\dif f}{\dif t}=\frac{\partial f}{\partial t}+\left [ f,{\cal H} \right ]=0,
\end{eqnarray}
where $[\cdot,\cdot]$ denotes a Poisson bracket. Here we confine ourselves to the linearized 
equation:
\begin{eqnarray}
\frac{\partial f_1}{\partial t}+\left [ f_1,{\cal H}_0 \right ]
+\left [ f_0,V_1 \right ]=-\left [ f_0,V_{\rm e} \right ].
\label{eq:linearized-Vlasov}
\end{eqnarray}
The remainder of this paper is devoted to the study of solutions of this
equation and their application to collisionless near-Keplerian discs. 

\section{The Finite-Element Method}
\label{sec:FEM-model}

The dynamics and stability of collisionless discs are usually studied
by one of two numerical methods: (i) $N$-body simulations
\citep[e.g.,][]{S87}; (ii) expansion of the perturbed gravitational
potential in a set of basis functions, followed by the evaluation of a
matrix representing the response of the disc to a given imposed
potential \citep[e.g.,][]{K77}. Neither of these methods, however, is ideal for
investigation of the oscillations and response of low-mass
near-Keplerian discs, for several reasons: (i) slow oscillations are
stable (T01) and therefore more difficult to detect
than growing modes; (ii) slow oscillations have low
frequencies, and thus $N$-body simulations must be followed for many
dynamical times; (iii) low-mass discs also support
short-wavelength fast (i.e., frequency independent of $\mu$)
oscillations and these cannot be resolved without a large set of basis
functions; (iv) we shall find that some slow oscillations have
nearly singular components. Here, we adopt a
finite-element method (FEM) and reduce the linearized CBE to a system
of ordinary differential equations that describes the temporal
evolution of the disc, both the eigenfrequency spectrum of an isolated
disc and the response of a disc to external perturbations. We use a
$C_0$ FEM (all functions are continuous, but not necessarily
differentiable at boundaries between elements) in the configuration 
space.  

In this section, we briefly review the principles of FEM modelling.  
For a general introduction see \cite{ZTZ05}. Detailed descriptions of the
application of an FEM to collisionless self-gravitating systems can be
found in \cite{J10} for perturbed systems and in \cite{JT10} for equilibrium 
models.

\subsection{Finite ring elements in the configuration space}

We split the configuration space into $N$ ring elements. The $n$th element 
is characterized by its nodes at $R_n$ and $R_{n+1}$, and by a linear
interpolating vector $\Gvec_n(R)$ defined by
\begin{eqnarray}
\Gvec_n=\left [
\begin{array}{cc}
    G_{1,n} & G_{2,n} 
\end{array}
\right ],~~G_{1,n}=1-\bar R,~~G_{2,n}=\bar R,
\label{eq:linear-interpolate}
\end{eqnarray}
where $\bar R=(R-R_n)/\Delta R_n$ and $\Delta R_n=R_{n+1}-R_n$.  Since we are
interested only in linear perturbations, disturbances of different azimuthal
wavenumber $m$ are independent. For the wavenumber $m$, the potential
$V_1$ and the surface density $\Sigma_1$ are thus computed from 
\begin{align}
V_1(R,\phi,t) =& {\rm Re} \sum_{n=1}^{N} 
H_n(R) \Gvec_n(R) \cdot \avec_n(t) e^{{\rm i}m\phi}, \\
\Sigma_1(R,\phi,t) =& {\rm Re} \sum_{n=1}^{N} 
H_n(R) \Gvec_n(R) \cdot \bvec_n(t) e^{{\rm i}m\phi}.
\end{align}
The function $H_n(R)$ is unity for $R_n\le R \le R_{n+1}$ 
and zero otherwise. The column vectors 
\begin{eqnarray}
\avec_n = \left [  
\begin{array}{cc}
a_{n1} & a_{n2}
\end{array}
\right ]^{\rm T},~~
\bvec_n = \left [  
\begin{array}{cc}
b_{n1} & b_{n2}
\end{array}
\right ]^{\rm T}, \nonumber
\end{eqnarray}
contain the nodal potentials and densities, respectively. According to the
definition of $\Gvec_n(R)$, $\Sigma_1$ is equal to ${\rm Re}\,
b_{n1}\exp(im\phi)$ at $R=R_{n}$ and to ${\rm Re}\,
b_{n2}\exp(im\phi)$ at $R=R_{n+1}$. Similarly, the nodal potentials at these
radii involve $a_{n1}$ and $a_{n2}$. The perturbed surface density and its
corresponding potential are continuous and differentiable inside 
elements and the continuity of these functions at the boundaries
of elements (nodes of rings) implies 
\begin{eqnarray}
a_{n2}=a_{n+1,1},~~ b_{n2}=b_{n+1,1}.
\label{eq:continuity-a-b-space}
\end{eqnarray}
This means that for a given $m$ we have $N_{\rm t}=N+1$ 
independent nodal potentials/densities. 

The angle-action representation of the perturbed potential $V_1$ reads
\begin{eqnarray}
V_1(\wvec,\Jvec,t)={\rm Re} \sum_{l=-\infty}^{+\infty} 
\tilde h_{1,l}(\Jvec,t) ~ e^{{\rm i}(l w_R+m w_{\phi} )},
\label{eq:V1-AA-space}
\end{eqnarray}
where 
\begin{eqnarray}
\tilde h_{1,l}(\Jvec,t) \!\!\! &=& \!\!\! \sum_{n=1}^{N}
{\bf \Psi}_{l}(n,\Jvec) \cdot \avec_n(t), \\
{\bf \Psi}_{l}(n,\Jvec) \!\!\! &=& \!\!\! 
\frac{1}{2\pi} \oint H_n(R) \Gvec_n ~ e^{{\rm i}m(\phi-w_{\phi})} 
e^{-{\rm i} l w_R} ~
\dif w_R.
\end{eqnarray}
The external disturbance $V_{\rm e}$ can also be expressed in terms of angle
and action variables.  To compute the perturbed DF $f_1(\wvec,\Jvec,t)$, we
use Fourier series of angle variables and write
\begin{eqnarray}
f_1(\wvec,\Jvec,t) = {\rm Re} \sum_{n=1}^{N}
\sum_{l=-\infty}^\infty \Evec_{l}(n,\Jvec) \cdot \zvec^n_{l}(t) ~
e^{{\rm i} (l w_R + m w_{\phi} )},
\label{eq:DF-summation-n-and-k}
\end{eqnarray}
where 
\begin{eqnarray}
\Evec_{l}(n,\Jvec)=\left [ 
\begin{array}{cc}
E_{l1}(n,\Jvec) & E_{l2}(n,\Jvec)
\end{array}
   \right ],
\end{eqnarray}
is an interpolating vector in the action space (to be specified in 
\S\ref{sec:interp}) and 
\begin{eqnarray}
\zvec^n_{l} = \left [ 
\begin{array}{cc}
z^n_{l1} & z^n_{l2}
\end{array}
   \right ]^{\rm T},
\end{eqnarray}
is a column vector of to-be-determined DFs whose elements should satisfy 
the continuity condition 
\begin{eqnarray}
z^n_{l2}=z^{(n+1)}_{l1}.
\end{eqnarray}
Equation (\ref{eq:DF-summation-n-and-k}) calculates the distribution 
of perturbed orbits based on their passage through ring elements in 
the configuration space. If an orbit stays only inside the $n$th element, 
its DF becomes
\begin{eqnarray}
\hat f_n(\wvec,\Jvec,t) = {\rm Re}
\sum_{l=-\infty}^{\infty} \Evec_{l}(n,\Jvec) \cdot \zvec^n_{l}(t) ~ 
e^{{\rm i} (l w_R+m w_{\phi})}.
\end{eqnarray}
In general, eccentric orbits may visit more than one ring element. The summation 
over $n$ in (\ref{eq:DF-summation-n-and-k}) takes this behavior into account.

\subsection{Projected evolutionary equations}
\label{sec:projected-evolutionary-equations}

We use the conditions (\ref{eq:continuity-a-b-space}) and assemble the 
nodal densities $\bvec_n(t)$ and potentials $\avec_n(t)$ in the global 
$N_{\rm t}$-dimensional vectors $\dvec(t)$ and $\pvec(t)$, respectively. 
Similarly, $\zvec^n_{l}(t)$ are collected in $\zvec_{l}(t)$. We now take the 
inner product of (\ref{eq:linearized-Vlasov}) with
\begin{eqnarray}
e^{ -{\rm i}( l'w_R + m w_{\phi} ) }
[\Evec_{l'}(n',\Jvec) ]^{\rm T},
\nonumber
\end{eqnarray}
and integrate the resulting systems of equations over the angle-action space
to obtain the Galerkin-weighted residual form of (\ref{eq:linearized-Vlasov})
as
\begin{eqnarray}
\Umat_1(l) \cdot \frac{\dif \zvec_{l}(t)}{\dif t} =
-{\rm i}\Umat_2(l)\cdot \zvec_{l}(t) + 
{\rm i} \Umat_3(l)\cdot \pvec(t) + {\rm i}\Zvec_l(t).
\label{eq:Galerkin-projection-Vlasov}
\end{eqnarray}
Here $\Umat_1$, $\Umat_2$ and $\Umat_3$ are constant square matrices 
of dimension $N_{\rm t}\times N_{\rm t}$, and $\Zvec_l(t)$ is an 
$N_{\rm t}$-dimensional column vector, which is the Galerkin projection 
of $-[f_0,V_{\rm e}]$.

One can also verify that the Galerkin projections of (\ref{eq:poisson-integral}) 
and (\ref{eq:density-DF-relation}) respectively become
\begin{eqnarray}
\pvec(t)=\Cmat \cdot \dvec(t),~~
\dvec(t)=\sum_{\lvec}\Fmat(l) \cdot \zvec_{l}(t).
\label{eq:p-and-d-vs-zvec}
\end{eqnarray}
The constant matrices $\Cmat$ and $\Fmat(l)$ are of dimension 
$N_{\rm t}\times N_{\rm t}$. We combine (\ref{eq:Galerkin-projection-Vlasov}) 
and (\ref{eq:p-and-d-vs-zvec}) to express $\pvec(t)$ in terms of 
$\zvec_{l}(t)$, and transform (\ref{eq:Galerkin-projection-Vlasov}) 
to a non-homogeneous ordinary differential equation for $\zvec_{l}(t)$:
\begin{eqnarray}
\frac{\dif \zvec_{l}(t)}{\dif t} \!\!\! &=& \!\!\!\
- {\rm i} \Umat^{-1}_1(l) \cdot \Umat_2(l)\cdot \zvec_{l}(t) + 
{\rm i} \Umat^{-1}_1(l) \cdot \Zvec_l(t) \nonumber \\
\!\!\! &+& \!\!\!  
\sum_{l'=-\infty}^{+\infty} {\rm i} \Umat^{-1}_1(l) \cdot 
\Umat_3(l) \cdot \Cmat \cdot \Fmat(l') \cdot \zvec_{l'}(t),
\label{eq:Galerkin-projection-Vlasov-vs-zvec}
\end{eqnarray}
for $l,l'=0,\pm 1,\pm 2,\cdots$. By defining 
\begin{eqnarray}
\zvec(t) = \left [
\begin{array}{ccccccc}
\ldots &
\zvec^{\rm T}_{-2} & \zvec^{\rm T}_{-1} &
\zvec^{\rm T}_{0}  & \zvec^{\rm T}_{+1} & 
\zvec^{\rm T}_{+2} & \ldots
\end{array}
   \right ]^{\rm T},
\end{eqnarray}
and collecting the elements of $\Umat^{-1}_1(l) \cdot \Zvec_l(t)$ 
(for all $l=0,\pm 1,\pm 2,\cdots$) in the global forcing vector 
$\Fvec(t)$, the system (\ref{eq:Galerkin-projection-Vlasov-vs-zvec}) 
can be cast into the standard form of linear evolutionary equations:
\begin{eqnarray}
\frac{\dif}{\dif t}\zvec(t)=-{\rm i} \Amat \cdot \zvec(t)+{\rm i} \Fvec(t).
\label{eq:forced-linear-equations}
\end{eqnarray}
In the absence of external disturbances, $\Fvec(t)=0$, 
the corresponding homogeneous equation admits a solution of the 
form $\zvec(t)=\exp(-{\rm i}\omega t)\cvec$ that yields the 
linear eigensystem:
\begin{eqnarray}
\Amat \cdot \cvec=\omega \cvec.
\label{eq:eigensystem}
\end{eqnarray}
We find the spectrum of $\omega$ using Hessenberg transformation 
of $\Amat$ followed by QR factorization. The eigenvector conjugate 
to a given eigenfrequency $\omega_j$ is then computed using the 
singular value decomposition 
\begin{eqnarray}
\Amat-\omega_j \Imat=\Vmat_1^{\rm T} \cdot \Wmat \cdot \Vmat_2,
\end{eqnarray}
where $\Wmat$ is a diagonal matrix whose elements are the singular 
values of $\Amat-\omega_j \Imat$, and $\Imat$ is the identity matrix. 
The column of $\Vmat_2$ corresponding to the smallest singular value 
is the eigenvector $\zvec^{(j)}$ associated with $\omega_j$.

\subsection{Interpolating functions in the action space}

\label{sec:interp}

In our $C_0$ FEM analysis, the local interpolating vector functions $\Gvec_n(R)$ 
(also known as shape functions) can reconstruct the spatial profile of any oscillatory wave 
whose wavelength is sufficiently large compared to the sizes of elements. 
However, we must also interpolate $f_1$ in the action space, which requires defining 
the interpolating vectors $\Evec_l(n,\Jvec)$ (eq.\ \ref{eq:DF-summation-n-and-k}). 
To do this we use arbitrary dynamic solutions of the linearized collisionless
Boltzmann equation, which should be an adequate representation of the DF for 
the purposes of interpolation. In the angle-action space, and using equation 
(\ref{eq:V1-AA-space}), one can show
\begin{eqnarray}
e^{{\rm i}m\phi}\Gvec_n(R) = \tilde V_1(n,\Jvec,\wvec) = \sum_{l=-\infty}^{+\infty} 
{\bf \Psi}_{l}(n,\Jvec) ~ e^{{\rm i} l w_R+{\rm i}m w_\phi}.
\label{eq:Gn-AA-space}
\end{eqnarray}
We assume $\partial f_1/\partial t=-{\rm i}\gamma f_1$, substitute 
$\tilde V_1(n,\Jvec,\wvec)$ into the linearized CBE and solve the resulting equation 
to obtain the interpolating vector in action space (cf.\ eq.\ \ref{eq:DF-summation-n-and-k})
\begin{eqnarray}
\Evec_{l}(n,\Jvec) =
\frac{l \partial f_0/\partial J_R+m\partial f_0/\partial J_{\phi} }
                        {l\Omega_R+m \Omega_{\phi}-\gamma} {\bf \Psi}_{l}(n,\Jvec). 
                        \label{eq:interpolating-DF-action-space}
\end{eqnarray}
The physical eigenfrequency $\omega$ will thus be equal to
$\omega_{c}+\gamma$ with $\omega_c$ being the computed eigenvalue of
(\ref{eq:eigensystem}). Varying $\gamma$
tests the robustness of our numerical methods since the results should
be independent of $\gamma$. Our tests show that in general our results
are insensitive to variations of $\gamma$. Nevertheless, the choice
$\gamma \gtrsim \max[\Omega_{\rm pr}(\Jvec)]$ offers better
performance, particularly in colder models. We originally used
$\gamma=0$, corresponding to the use of static CBE solutions as
interpolating vectors in action space, but with this choice we found
occasional spurious growing modes.

We remark that we do not generate a finite-element mesh in action space, 
for two reasons: (i) to reduce the size of the Galerkin-weighted evolutionary 
equations; (ii) to avoid creating spurious growing modes. The second of these 
properties has a straightforward mathematical explanation: the number 
of reachable eigenmodes in the
configuration space is equal to the number of independent nodal
variables, which is $N+1$ in our FEM analysis. However, the eigenvalue
problem (\ref{eq:eigensystem}) has been formulated in the phase space
and the number of computed eigenmodes is equal to $(N+1)\times (l_{\rm
  max}- l_{\rm min}+1)$ where $l_{\rm min}<0$ and $l_{\rm max}>0$ are
the lower and upper bounds in the $l$ sums. Since the nodal densities
$\dvec$ are related to $\zvec_l$ through equation
(\ref{eq:p-and-d-vs-zvec}), there will be $(N+1)\times (l_{\rm max}-
l_{\rm min})$ computed eigenmodes more than $N+1$ eigenmodes that the
dimension of $\dvec$ determines. The extra modes should therefore
overlap in groups of $(l_{\rm max}- l_{\rm min})$ members to avoid
spurious modes. This happens in our numerical calculations performed
in \S\ref{sec:prograde-waves} when the Fourier expansions over $w_R$
converge inside all ring elements, as is expected for the
reconstruction of $V_1$ and $f_1$ in the $(w_r,w_{\phi})$-subspace.
Generating a finite-element mesh, let us say with $N_a$ nodes in the
two-dimensional $\Jvec$-space, will result in $2N_a\times (l_{\rm
  max}- l_{\rm min}+1)$ modes, but assuming the convergence of Fourier
series, only $(N+1)$ groups of them will correspond to eigenmodes in
the configuration space. Consequently, $2N_a-N+1$ computed modes will
be spurious, and our calculations show that such spurious modes are
growing. Working with $2N_a=N+1$ will not help because it does not
necessarily guarantee the convergence of FEM model in the action
space.  

Only few modes out of $N+1$ possible states in the configuration space
(see \S\ref{sec:prograde-waves}) are physical. The rest are either
singular, or do not satisfy the boundary conditions as $R\rightarrow
\infty$.  Note that for frequencies $\omega$ that lie between the
maximum and minimum of the precession frequency $\Omega_{\rm pr}$ the
singular modes may be van Kampen modes (restricted to the surface in
action space on which $\omega=\Omega_{\rm pr}$) which would be damped
by the Landau mechanism. However, not all modes with frequency in this
range are necessarily van Kampen modes, since the mode may not be
produced by the orbits associated with that resonance. Thus discrete
modes may overlap in frequency space with continuous modes. 

\section{Prograde waves}
\label{sec:prograde-waves}

Our finite-element mesh is uniform in log radius, 
\begin{align}
R_n =&10^{-\alpha_1+\alpha_2 y(n,N)}, \\
y(n,N) =& \frac{1}{2(N+1)}+\frac{n-1}{N+1},~~n=1,2,\cdots,N.
\label{eq:grid}
\end{align}
The numerical computation of the Fourier coefficients ${\bf \Psi}_{\lvec}(n,\Jvec)$ 
and then the interpolating vectors $\Evec_{\lvec}(n,\Jvec)$ needs a mesh in the 
$(a,e)$ space. Such a grid is not arbitrary because there must be at least one orbit 
that visits the $n$th ring element in configuration space.  We fulfill this requirement 
using the following two-dimensional grid
\begin{eqnarray}
(a_i,e_j) = \left [ R_i,y(j,M_{e})\right ], \nonumber
\end{eqnarray}
where the grid points along the $a$-direction exactly coincide with the 
boundary nodes of the mesh in the configuration space, and there are 
$j=1,2,\cdots,M_e+1$ grid points in the $e$-direction. A circular orbit 
is indeed assigned to each boundary of a ring element. This is particularly 
helpful in cold discs where one must interpolate the population of circular 
orbits. The parameters $\alpha_1$ and $\alpha_2$ are chosen so that the 
computed disc mass $4\pi^2 \int f(\Jvec) ~ \dif \Jvec$ using the grid 
points in the $(a,e)$-space agrees with the actual disc mass within 1\%. 

In this paper we focus on slow modes with azimuthal wavenumber $m=1$. 
Slow modes exist with larger $m$, so long as the calculation includes Fourier 
terms with index $l=-m$. In particular, we have found a number of isolated, 
non-singular $m=2$ modes; these are present only if we use a fine FEM mesh, 
since they are more compact and have shorter wavelengths than the $m=1$
modes. The wavelengths of $m=2$ modes shrink to zero as the disc
becomes colder (see Appendix). 
This behavior is expected since the only large-scale slow modes in cold low-mass 
discs have $m=1$. We found no unstable modes, which is also expected for 
low-mass discs (T01).

\begin{figure}
\centerline{\hbox{\includegraphics[width=0.45\textwidth]{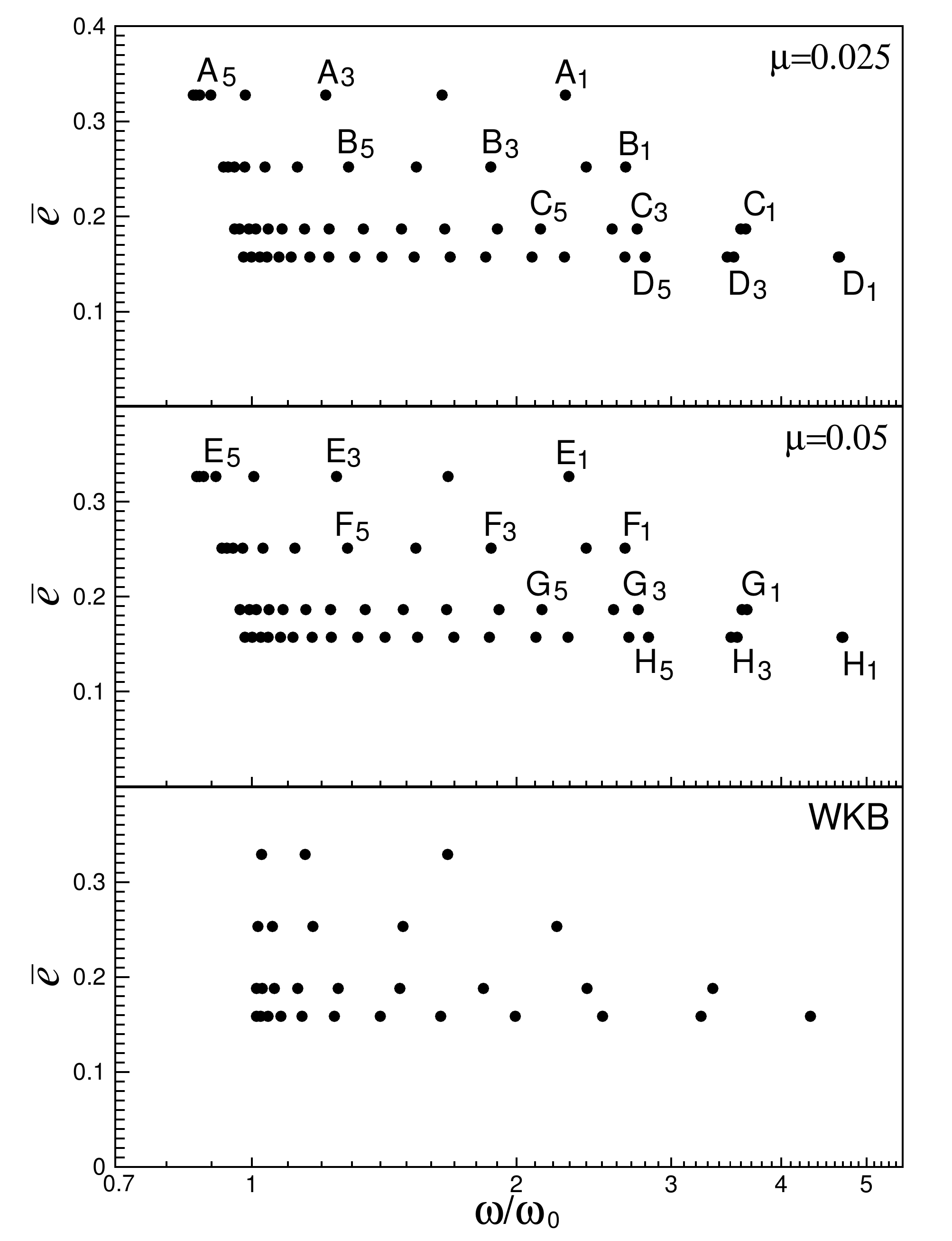}} }
\caption{Eigenfrequency spectra of stable, prograde density waves in
  near-Keplerian discs. Only parent modes are shown. The vertical axis 
  is the mean eccentricity $\bar e$ and the horizontal axis is $\bar\omega=\omega/\omega_0$
  where $\omega_0=0.05861\mu+{\cal O}(\mu^2)$. These models correspond
  to DFs of the form (\ref{eq:equilibrium-DF}) with
  $K=5,10,20,29$. The calculation includes Fourier terms
  $l=-2,-1,0,1$. Note the logarithmic scale of the horizontal
  axis. {\it Top}: $\mu=0.025$ and $\omega_0=0.00146$.  {\it Middle}: $\mu=0.05$ and
  $\omega_0=0.00293$. The eigenfrequencies of modes ${\rm D}_1$ and
  ${\rm D}_2$ are very close and indistinguishable in the plots.  They
  are $\bar\omega_{{\rm D}_1}=4.659$ and $\bar\omega_{{\rm
      D}_2}=4.647$.  Similarly, we have $\bar\omega_{{\rm H}_1}=4.700$
  and $\bar\omega_{{\rm H}_2}=4.689$.  Note the similarity of the
  spectra in the top and middle diagrams despite the change of a factor
  of two in the disc mass $\mu$; this feature is characteristic of
  slow modes.  Mode shapes associated with the labelled frequencies
  have been plotted in Figures \ref{fig5}, \ref{fig6} and
  \ref{fig7}. {\it Bottom}: Eigenfrequency spectra derived from the WKB
  approximation described in the Appendix. Each plotted point represents a
  degenerate leading/trailing pair of modes.}
\label{fig4}
\end{figure}

We began our calculations with $N=M_{e}=70$ and $l=-1$, and
increased the number of Fourier terms and ring elements until the
eigenfrequencies of stable modes found from (\ref{eq:eigensystem})
converged to a fractional accuracy of $10^{-4}$. Typically this
required computing all Fourier terms with $-2\le l\le 3$ and a grid
with $N=160$ and $M_e=140$ ($N=180$ and $M_e=140$ for the models with
the lowest rms eccentricity, corresponding to $K=29$). We have also
experimented with including terms with larger values of $| l |$ but
these had only a small effect on our results. Taking grid
points in the regions with tiny values of $f(a,e)$ (see Figure
\ref{fig2}) leads to large errors in the properties of the calculated
density waves because the FEM discretization errors become larger than
the absolute magnitudes of physical quantities. We evade this
difficulty by generating the FEM mesh only in the annular region $0.01 \le
R \le 100$ using the parameters $(\alpha_1,\alpha_2)=(2,4)$ in
equation (\ref{eq:grid}). 

All non-singular eigenmodes with $m=1$ were found to be prograde
($\omega>0$). We find two general types of modes: a parent family that
is already present when only the $l=-1$ Fourier component is included
in the calculation, and a child family that bifurcates from the parent
family as more $l$-terms are included. The eigenfrequencies of child
modes are very close to those of their parent mode (typically within 1--2\%).
They emerge from resonant interactions between two approximate modes that are weakly 
coupled: the parent modes and singular van Kampen modes. For $l=0$ and $l=+1$ 
the singular components of the child modes correspond to the corotation (CR) and 
outer Lindblad (OLR) resonances, respectively. The coupling between slow and 
van Kampen modes is probably due mostly to highly eccentric orbits
that are perturbed by the gravity from both waveforms. The main evidence 
for this is that as the mean eccentricity $\bar e$ shrinks, child modes collapse to 
singular modes and disappear. 

We denote the maximum precession rate of circular orbits by 
$\omega_{0}={\rm max}[\Omega_{\rm pr}(\Jvec)]$; from equation
(\ref{eq:precrate}) $\omega_0=0.05861\mu+{\cal O}(\mu^2)$. We then
plot our results using the normalized frequency $\bar\omega=\omega/\omega_0$. 
Figure \ref{fig4} shows the eigenfrequency spectra of prograde $m=1$ 
parent modes for the mass ratios $\mu=0.025$ and $\mu=0.05$, and for 
four values of the mean eccentricity $\bar e$. The frequencies of child modes 
are not shown to avoid overcrowding the diagrams. Although the maximum 
precession rate $\omega_0$ is proportional to $\mu$, the spectra of 
$\bar\omega$ agree to within 1\% in the models with $\mu=0.025$ and $\mu=0.05$.  
This scaling shows that our results can be directly applied to all discs with mass ratios 
$\mu\ll1$, in particular to the tiny mass ratios $\mu \lesssim {\cal O}(10^{-3})$
characteristic of debris discs. 

\begin{figure*}
\centerline{\hbox{\includegraphics[width=0.45\textwidth]{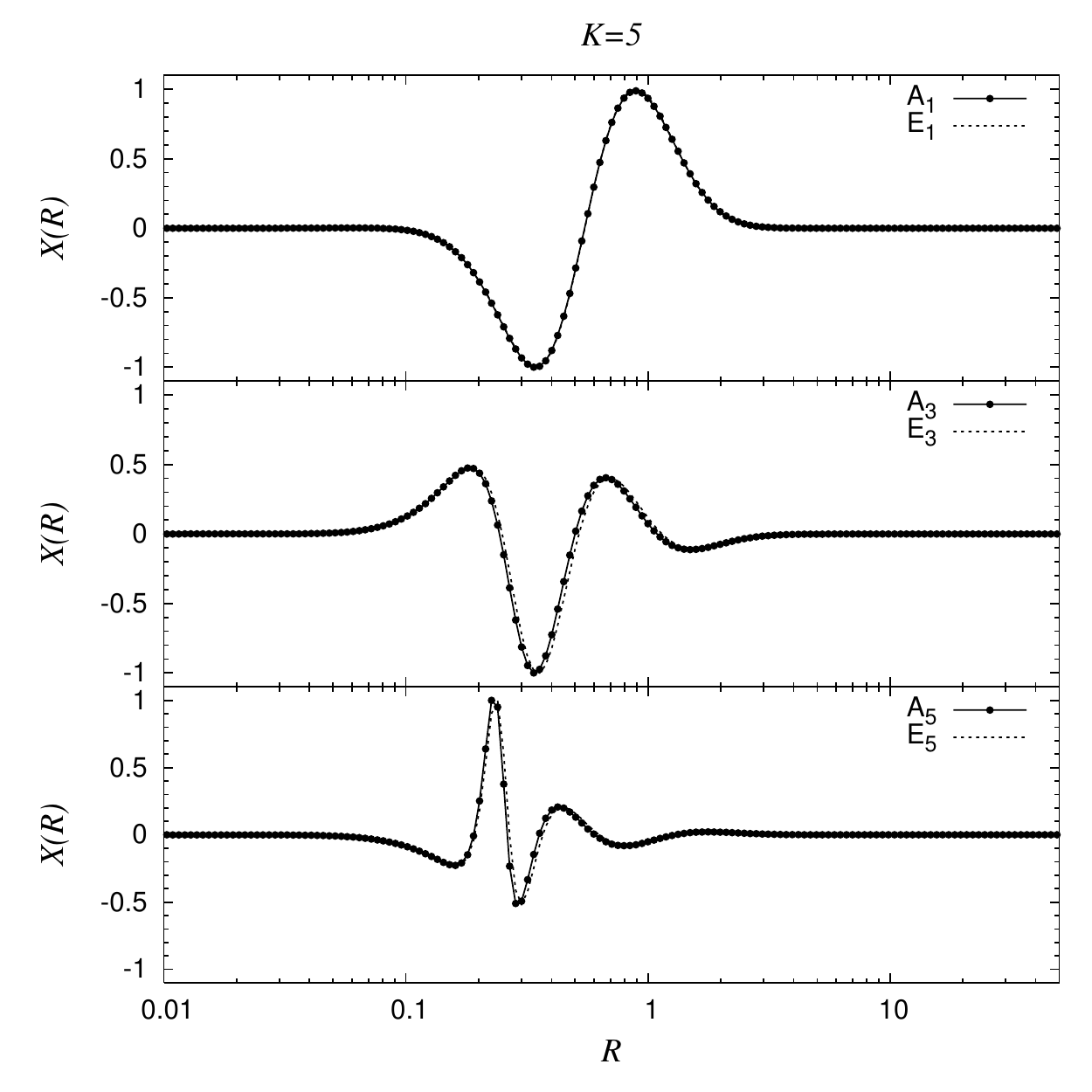}}
\hspace{0.1in}
             \hbox{\includegraphics[width=0.45\textwidth]{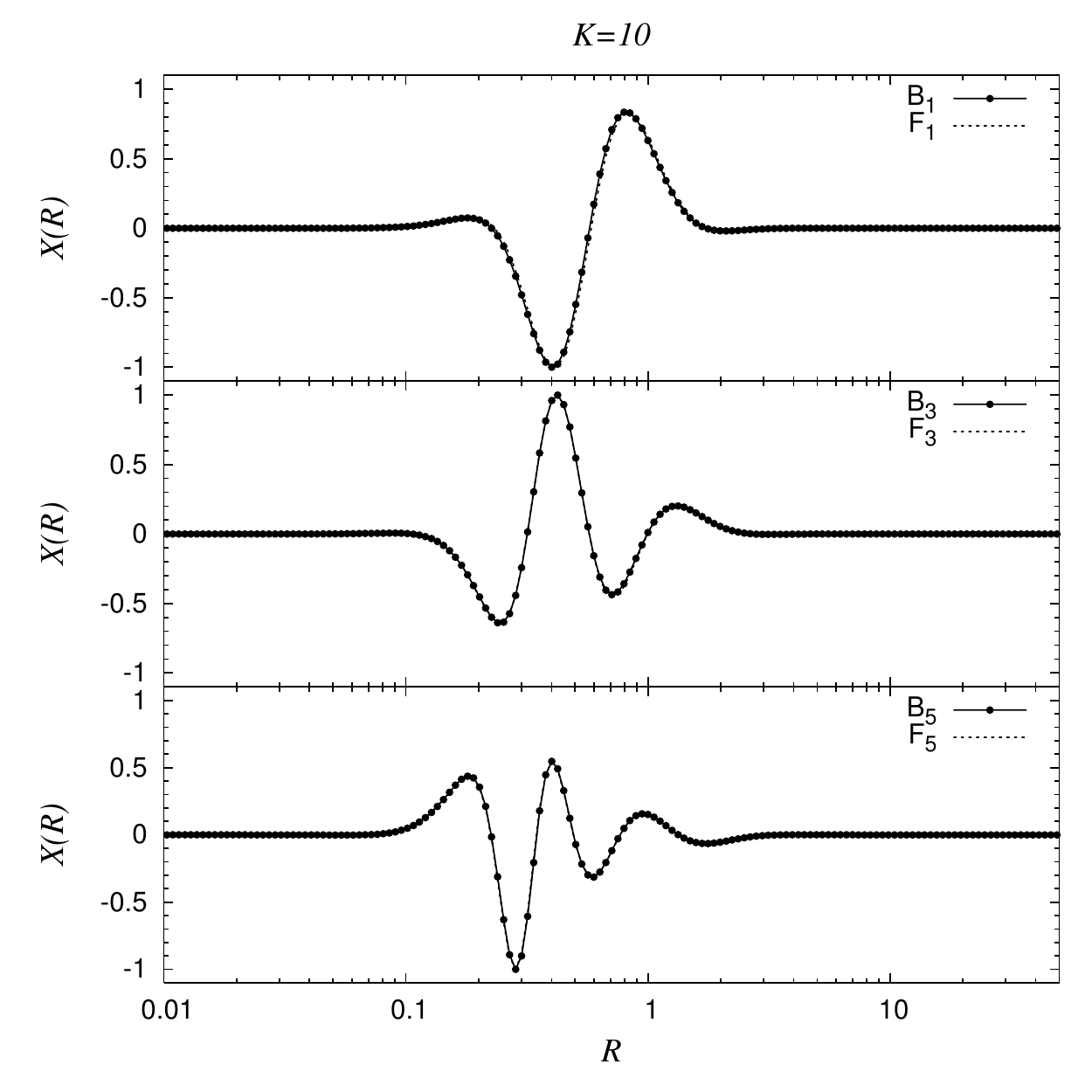}}
	    }	
\centerline{\hbox{\includegraphics[width=0.45\textwidth]{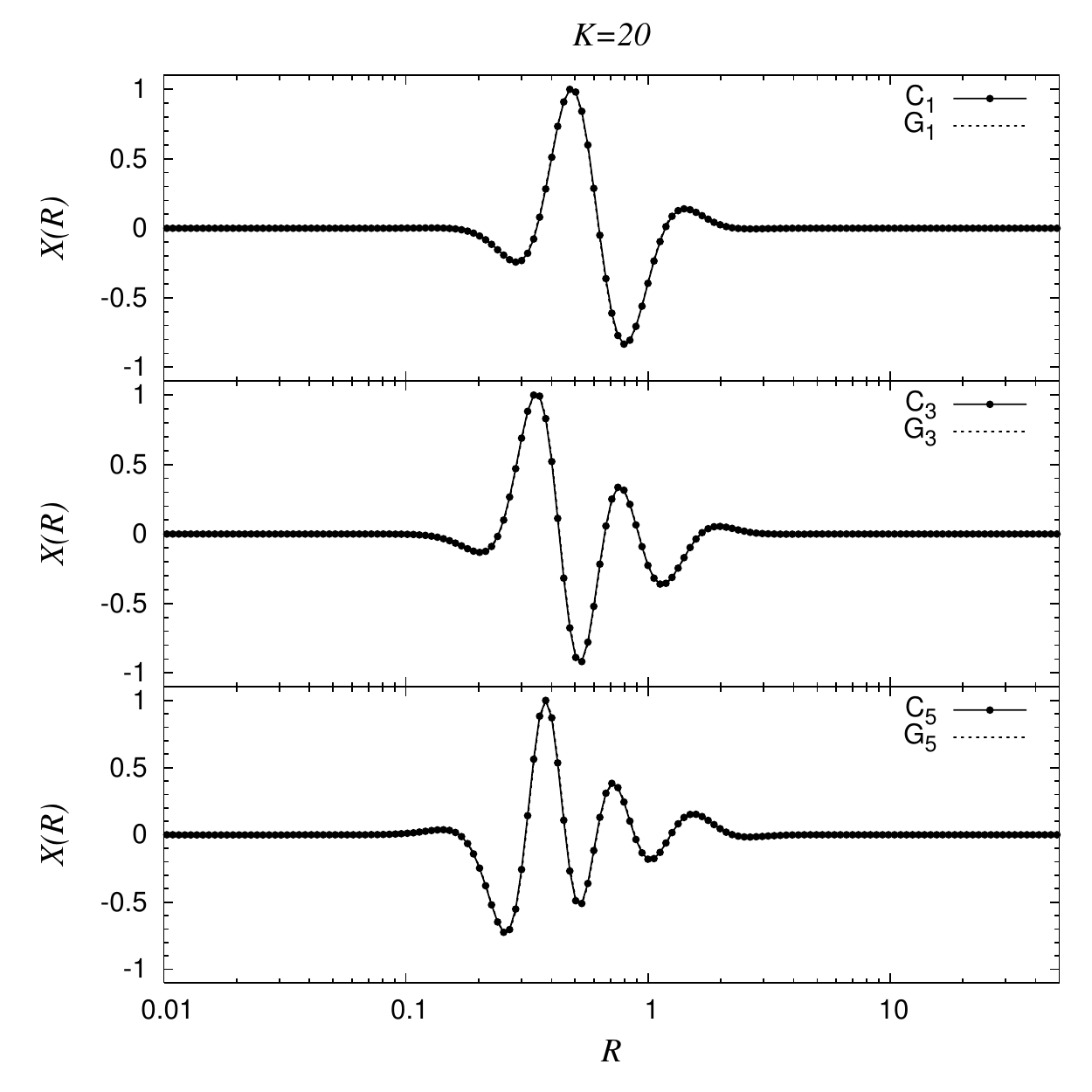}}
\hspace{0.1in}
             \hbox{\includegraphics[width=0.45\textwidth]{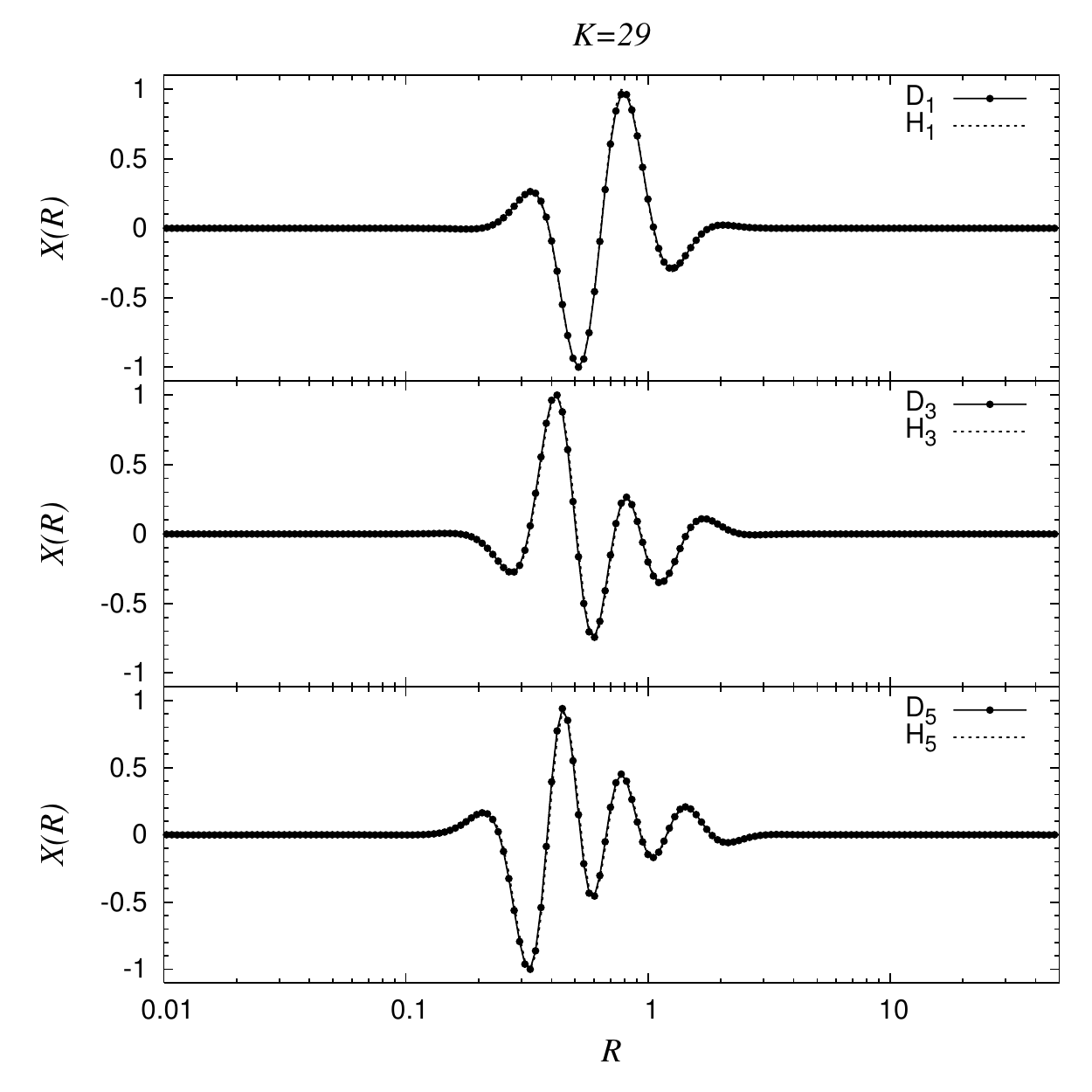}}
	    }	  	        
\caption{Perturbed density components $X(R)$ (cf. eq.\ \ref{eq:xy})
for some stable modes in near-Keplerian discs of $\mu=0.025$ (solid lines) and 
$\mu=0.05$ (dotted lines). In several panels the dotted line is not visible
because it lies under the solid line. There are $N=160$ ring elements in the 
configuration space for $K=5, 10$ and 20, and $N=180$ elements for $K=29$. 
Filled circles mark the locations of element nodes. In all panels, the maximum 
of $|X(R)|$ has been normalized to unity.}
\label{fig5}
\end{figure*}

Figure \ref{fig4} shows that the modes become more closely spaced as
their frequency decreases and the minimum frequency in each spectrum
is an accumulation point. This implies the existence of prograde waves
with arbitrarily short wavelengths. There is also a nice correlation
between the precession rate of the most eccentric orbit in the model
(see Figures \ref{fig1} and \ref{fig2}) and the lowest frequency in
the spectrum. Models with highly eccentric orbits have an accumulation
point of lower frequency. Figure \ref{fig4} shows that the number of
modes increases with decreasing $\bar e$. In the limit of $\bar
e\rightarrow 0$, however, dispersion-supported waves (or $p$-modes)
can not exist according to WKB results (T01). 

The frequency spacing between modes
$B_1$ and $B_2$ is larger than the spacing between $C_1$ and $C_2$, which 
in turn is larger than the spacing between $D_1$ and $D_2$ (which is so small 
that the two points are indistinguishable in the Figure). Similar behavior is seen 
in the $F$, $G$, and $H$ families in the middle panel of the Figure. In the limit 
$\bar e\rightarrow 0$, the parent modes tagged with the numbers $2j+1$ and $2j+2$ 
($j=0,1,2,\cdots$) become degenerate. In the language of T01, they form a degenerate 
leading/trailing pair of $p$-modes (see Appendix). The pairing process begins from 
modes with highest pattern speeds, for the resonant cavities of those modes are fed 
mostly by near-circular orbits, which are the only population used in WKB analysis. 
The child modes of degenerate pairs also disappear because their
supporting eccentric orbits disappear as $\bar e\rightarrow 0$. Modes with $\bar\omega\rightarrow 1$ 
and sufficiently large $\bar e$ engage highly eccentric orbits and thus lead to more complex dynamics.
Eccentric orbits are indeed the backbones of discs, and when perturbed, they affect a 
vast radial domain while near-circular orbits have only a local influence on developing 
patterns. 

We now examine the shapes of the modes. After finding $\omega$, we calculate its corresponding eigenvector
$\cvec$, and use this to compute the nodal potentials $\pvec$ and nodal 
densities $\dvec$ from (\ref{eq:p-and-d-vs-zvec}). Defining 
\begin{eqnarray}
X(R) \!\!\! &=& \!\!\! {\rm Re} \sum_{n=1}^{N} 
H_n(R) \Gvec_n(R) \cdot \bvec_n, \\
Y(R) \!\!\! &=& \!\!\! {\rm Im} \sum_{n=1}^{N} 
H_n(R) \Gvec_n(R) \cdot \bvec_n,
\end{eqnarray}
one can compute the perturbed density patterns
\begin{eqnarray}
\Sigma_1(R,\phi,t) = X(R) \cos (m\phi-\omega t) - 
Y(R)\sin(m\phi-\omega t),
\label{eq:xy}
\end{eqnarray}
for a single wavenumber $m$.  Note that $\bvec_n$ are extracted from
the elements of $\dvec$ using the following formula:
\begin{eqnarray}
\bvec_n=\left [
\begin{array}{cc}
d_{n} & d_{n+1}
\end{array}
\right ]^{\rm T},~~ n=1,2,\cdots,N. 
\label{eq:find-b-from-d}
\end{eqnarray}

Figure \ref{fig5} shows the profile of $X(R)$ for the labelled parent modes of
Figure \ref{fig4}.  Not only are the normalized frequencies of the modes
$A_j$ and $E_j$ identical, but also their mode shapes are very
similar. These remarks apply to the pairs $(B_j,F_j)$, $(C_j,G_j)$ and
$(D_j,H_j)$ as well, and demonstrate that the waveforms are independent of 
$\mu$ so long as $\mu\ll1$, as one would expect for slow modes. The figure 
also shows that the wavelength of oscillations increases with the pattern speed
$\omega$ in a given disc, and decreases as the mean eccentricity of the disc 
shrinks. The number of nodes increases as the frequency decreases. 
An interesting property of the waves showing multiple nodes is that their
density peaks are approximately equally spaced in logarithmic scales.

\begin{figure}
\centerline{\hbox{\includegraphics[width=0.45\textwidth]{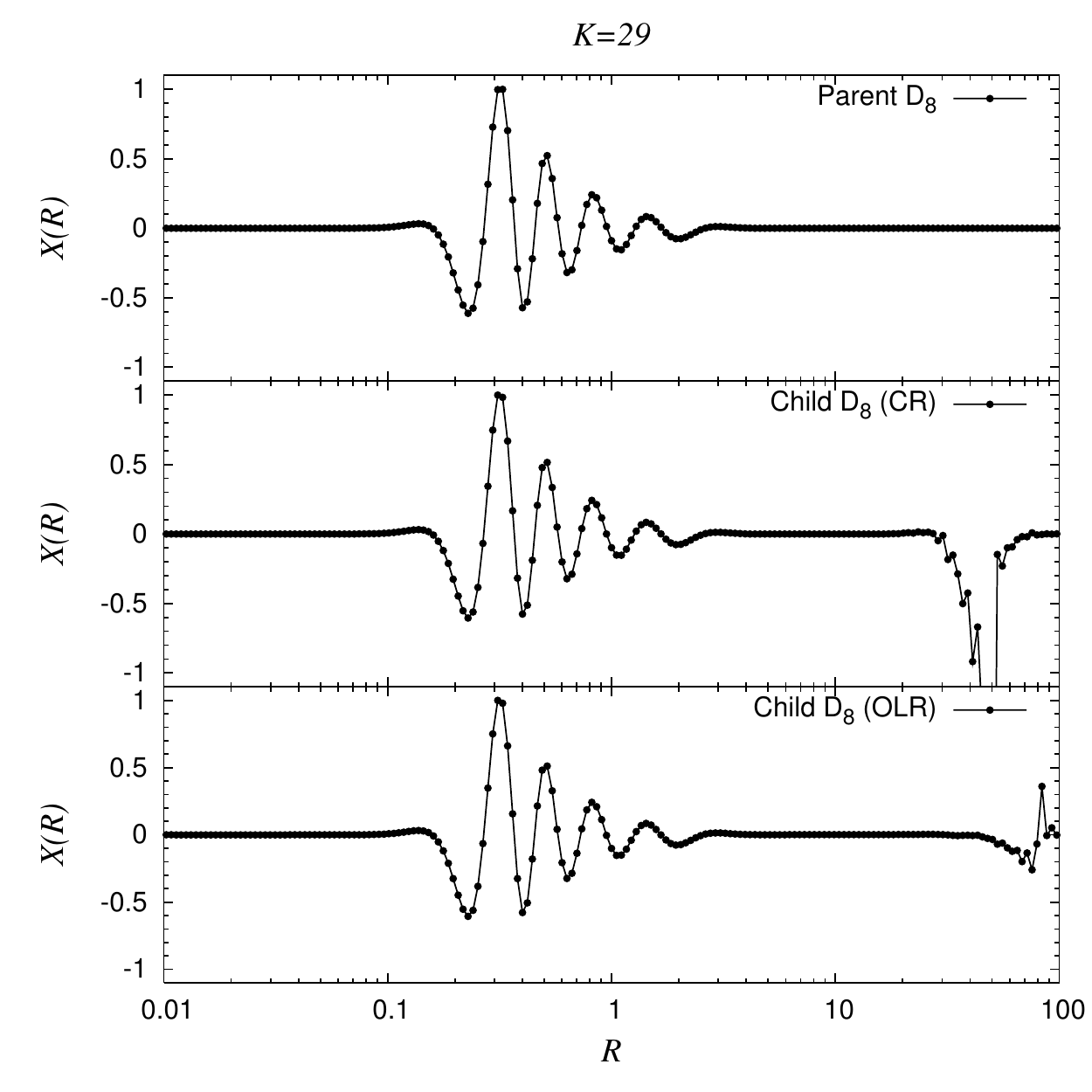}} }
\caption{Perturbed density components $X(R)$ for the parent mode ${\rm D}_8$
and two of its associated child modes ${\rm D}_{8,{\rm CR}}$ and ${\rm D}_{8,{\rm OLR}}$, 
in which the parent mode is coupled to singular modes at the corotation (CR) and outer Lindblad (OLR) resonances. 
The model parameters are $\mu=0.025$ and $K=29$. }
\label{fig6}
\end{figure}

The child modes are hybrid modes that inherit the features of their parents 
in the central regions of the disc, but have a spike at the location
of singular modes that couple to them. 
Figure \ref{fig6} displays the parent mode ${\rm D}_8$ of frequency $\bar\omega=2.083$
(see Figure \ref{fig4}) and its children ${\rm D}_{8,{\rm CR}}$ with $\bar\omega=2.0765$ 
and ${\rm D}_{8,{\rm OLR}}$ with $\bar\omega=2.0728$, which contain 
singular van Kampen modes at the corotation and outer Lindblad
resonances, respectively. In low-mass discs, these resonances are at
large radii where the surface density is small, so the singular
component of a child mode involves only a small fraction of the mass
involved in the parent mode. As the disc mass shrinks to zero the child 
modes merge with the parent mode. The reason is that the eigenfrequency 
of the parent mode is proportional to the disc mass so with very small disc 
masses the corotation and outer Lindblad resonances are at extremely large 
radii where the surface density is negligible. Thus the distinction between
parent and child modes is unimportant for low-mass discs such as debris discs.

\begin{figure*}
\centerline{\hbox{\includegraphics[width=0.33\textwidth]{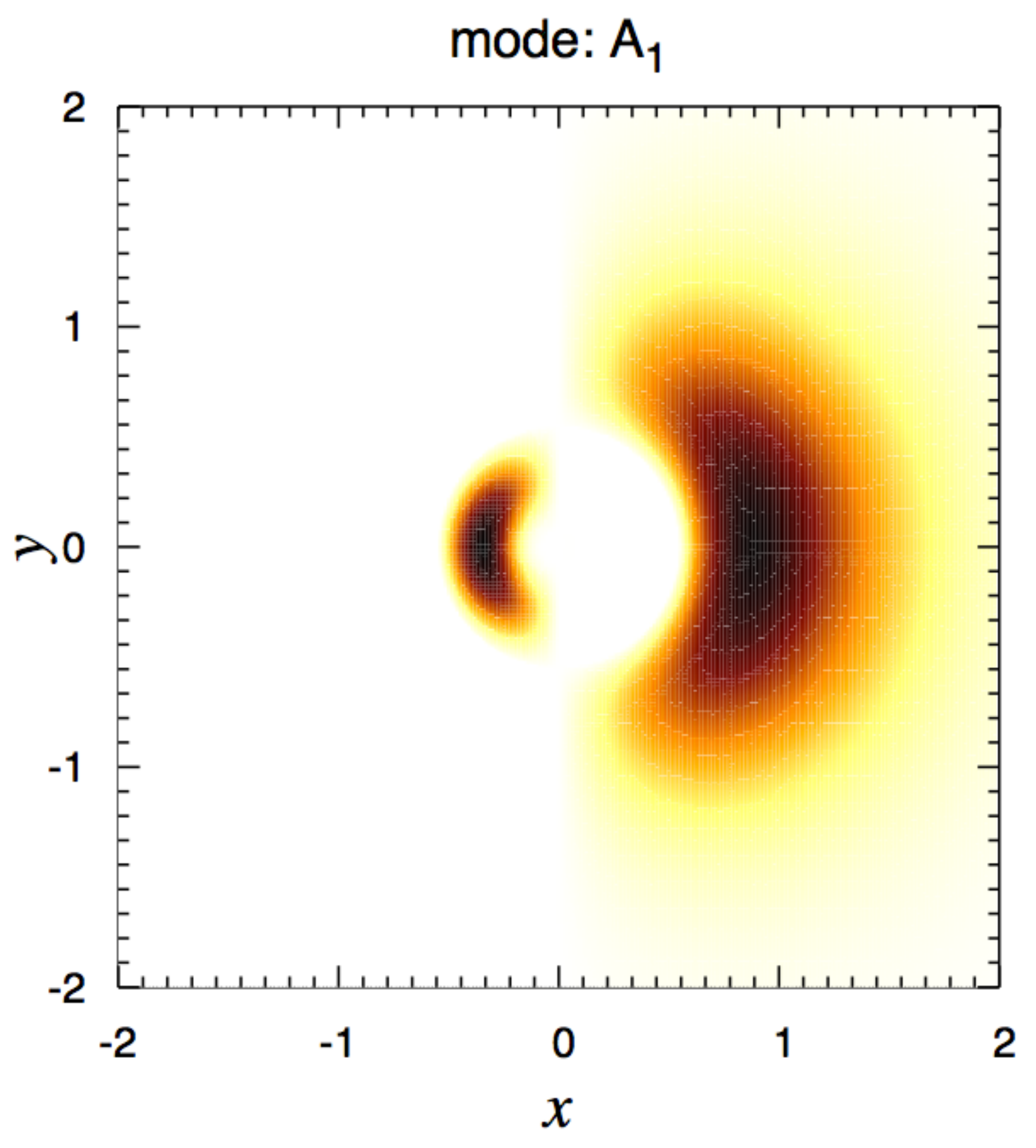}}
                    \hbox{\includegraphics[width=0.33\textwidth]{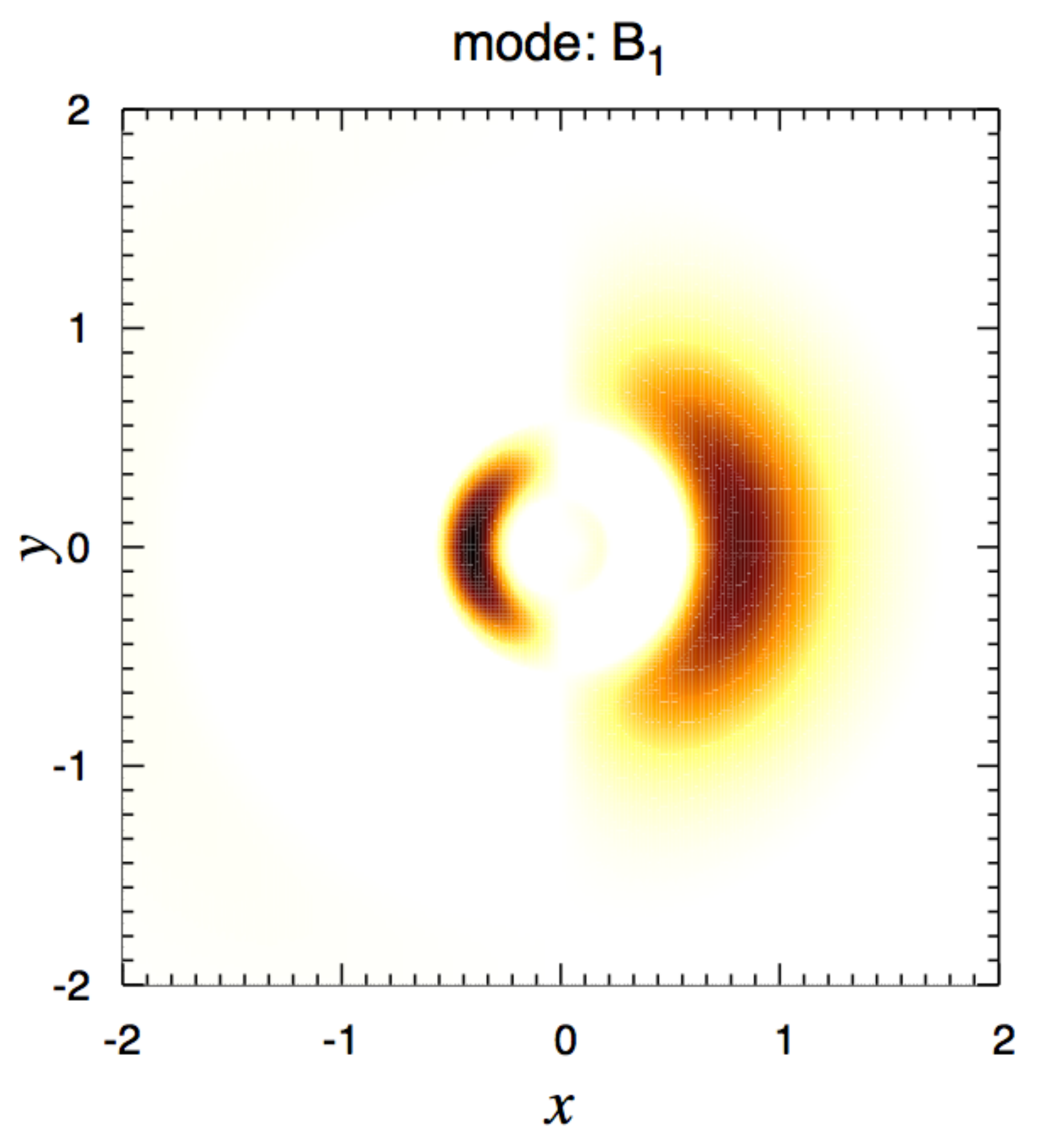}}            
                    \hbox{\includegraphics[width=0.33\textwidth]{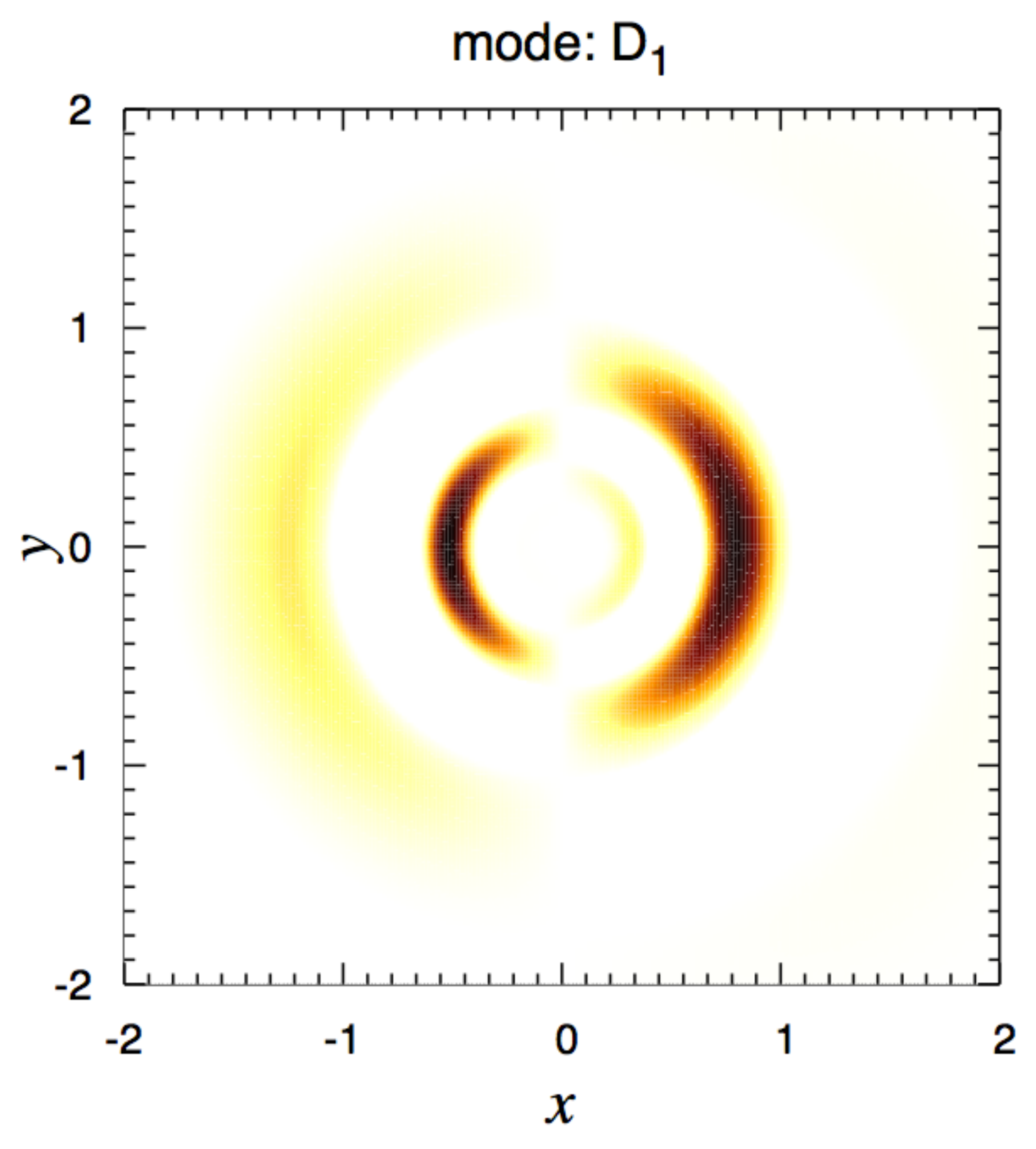}}
	          }
\caption{The patterns of oscillatory waves in the configuration 
space for a near-Keplerian disc with the mass ratio $\mu=0.025$. 
We have displayed only the positive part of $\Sigma_1(R,\phi,t)$ at 
$t=0$. Maximum densities of all panels have been normalized to unity, 
and the contour levels range from 0 to 1. The point mass sits at $(0,0)$.
{\it Left}: $K=5$. {\it Middle}: $K=10$. {\it Right}: $K=29$.}
\label{fig7}
\end{figure*}

Figure \ref{fig7} displays shaded contour plots of the pattern of $\Sigma_1(R,\phi,t)$ for some models with 
$\mu=0.025$ (mode shapes corresponding to $\mu=0.05$ are similar). It is seen that 
the wave packets are more radially compact in the colder ($K=29$) model than 
warmer ($K=5,10$) ones. 

The properties of these modes can be explored using the short-wavelength 
or WKB approximation described in the Appendix. The validity of this approximation 
requires $k >h/R$ where $k$ is the wavenumber and $h$ is 
a dimensionless number of order unity. If two adjacent nodes of a wave are at $R_1$ 
and $R_2$ then $\int_{R_1}^{R_2}k \,dR=\pi$ so the condition for validity 
of the WKB approximation may be written $\pi > h\log R_2/R_1$ or $\log_{10}R_2/R_1<1.36/h$. 
Inspection of Figure \ref{fig6} shows that for $h=1$ this condition is satisfied by all of 
the modes we have computed, though not by much in some cases. 

The bottom panel of Figure \ref{fig4} shows the WKB frequency
spectrum. Each point corresponds to a degenerate pair of modes, one
composed of leading spiral waves and the other of trailing. These
modes arise from waves in the resonant cavities defined by the closed
frequency contours in Figure \ref{fig:cont}. The WKB approximation
correctly reproduces several striking features of the FEM frequency
spectra: (i) all modes are prograde ($\omega>0$); (ii) both the number
of modes and the maximum frequency grow as the mean eccentricity $\bar
e$ of the disc shrinks; (iii) there is an accumulation point of modes
near $\omega/\omega_0=1$ in the FEM spectra and at $\omega/\omega_0=1$
in the WKB spectra; (iv) there is also reasonable quantitative
agreement between the frequencies derived by the two methods, at least
for the discs with the lowest mean eccentricity. The WKB analysis in
the Appendix fails to find the child modes, for two reasons: (i) it is
based on the epicycle approximation, which assumes that the
eccentricity is small and thus neglects the highly eccentric orbits
that couple the slow and van Kampen modes; (ii) it is based on the
approximation that the disc mass $\mu\rightarrow 0$, and in this limit
the pattern speed of the slow mode goes to zero so the outer Lindblad
and corotation resonances are at very large radii where the disc
surface density is negligible.

\section{Excitation of oscillatory waves}

\label{sec:excite}

Protostars live in the harsh environments of their birth
clusters. Simulations of the Orion nebula \citep{SC01} show that about
10 per cent of stars can have encounters closer than 100 AU within
$10^7$ years. Such encounters can excite waves in planetesimal/debris
discs. Encounters were invoked as a possible explanation for the
asymmetries in the $\beta$ Pictoris debris disc by \cite{kj95} and
\cite{lk01} but these authors treated the debris disc as a collection
of test particles, which can give misleading results since the
self-gravity of the disc dominates the apsidal precession. 

Since our goal here is only to illustrate this process we confine ourselves to 
in-plane parabolic encounters. Consider a disc particle orbiting around a star 
of mass $M_{\star}$, and assume a perturber of mass $M_p$. As in earlier sections, 
we scale all lengths so that the disc length scale $b$ is unity, and denote the 
normalized position vectors of the particle and perturber (with respect to the host star) 
by $\Rvec$ and $\Rvec_p$, respectively. The equation of motion for a disc particle is
\begin{eqnarray}
\frac{\dif ^2\Rvec}{\dif t^2}=-\bnabla \left [ \avec_{\star}\cdot \Rvec + V_0(\Rvec) + V_1(\Rvec,t) + 
V_{p}(\Rvec,\Rvec_p) \right ], 
\end{eqnarray}
where $\avec_{\star}$ is the acceleration vector of the host star in an inertial frame, $V_0$ is 
the unperturbed gravitational potential due to $M_{\star}$ and the self-gravity of the disc, 
$V_1$ is the perturbed self-gravitational potential of the disc, and $V_{p}=-(M_p/M_{\star})/|\Rvec_p-\Rvec|$ 
is the potential field of the perturber. The gradient $\bnabla$ is taken over the $\Rvec$ space,
and the normalized time $t$ is related to the actual time $t_{\rm actual}$ through 
$t/t_{\rm actual}=(G M_{\star} /b^3)^{1/2}$.

We assume that $\avec_{\star}$ is due to the encounter; thus we ignore the cluster's tidal field. 
Consequently,
\begin{eqnarray}
\avec_{\star}\cdot \Rvec=\left ( \frac{M_{\rm p} }{ M_\star} \right ) \frac {\Rvec_p \cdot \Rvec}{R_p^3},
~~ R_p=|\Rvec_p|.
\end{eqnarray}
For a distant encounter, $R\ll R_p$, the potential $V_p$ can be 
expanded as the following series
\begin{eqnarray}
\!\!\! &{}& \!\!\! V_p = -\left ( \frac{M_p}{M_\star} \right ) \frac{1}{R_p} \sum_{i=0}^{\infty} 
\left ( \frac {R}{R_p} \right )^i
P_i \left [ \cos \left ( \phi-\phi_p \right ) \right ], 
\label{Vp-expansion} \\
\!\!\! &{}& \!\!\! \cos \left ( \phi-\phi_p \right ) =
\frac{\Rvec_p \cdot \Rvec}{R_p R}, \nonumber
\end{eqnarray}
where $\phi$ and $\phi_p$ are, respectively, the azimuths of 
the disc particle and perturber measured from an inertial reference
line, and $P_i$ are Legendre polynomials. The effective potential 
due to flying-by perturber thus reads
\begin{eqnarray}
V_{\rm e} \!\!\! &=& \!\!\! \avec_{\star} \cdot \Rvec+V_p, 
\nonumber \\
\!\!\! &=& \!\!\! -\left ( \frac{M_p}{M_\star} \right ) \frac{1}{R_p} \sum_{i=2}^{\infty} 
\left ( \frac {R}{R_p} \right )^i
P_i\left [ \cos \left ( \phi-\phi_p \right ) \right ],
\label{eq:Ve-expansion}
\end{eqnarray}
where we have dropped the $i=0$ term in (\ref{Vp-expansion}) 
because it makes no contribution to the force, and the 
$i=1$ term has been cancelled by $\avec_{\star} \cdot \Rvec$. 

Modes having azimuthal wavenumber $m=1$ can only be excited by those 
$i\ge 3$ terms of $V_{\rm e}$ that produce $\cos \phi$ and $\sin \phi$ factors. 
Modes with $m=2$ are excited by the $i=2$ term of $V_p$, which is much larger 
than the $i \ge 3$ terms for distant perturbers ($R/R_p\ll1$); however, we have 
found (\S\ref{sec:prograde-waves}) that slow modes with $m=2$ have wavelengths 
that are generally smaller than those of $m=1$ modes, even for discs with a 
relatively large mean eccentricity $\bar e$, and which shrink to zero as $\bar e \rightarrow 0$. 
Thus $m=2$ modes couple less effectively to smooth perturbing potentials. 
We conclude that the dominant slow mode excited by an external perturber may 
have either $m=1$ or $m=2$.

For brevity, we shall examine only $m=1$ modes here. The dominant term of (\ref{eq:Ve-expansion})
for $m=1$ perturbations is
\begin{eqnarray}
V_{\rm e} \simeq  -\frac{3}{8} \left (  \frac{M_p}{M_\star} \right )
                                 \left ( \frac{1}{R_p} \right ) \left ( \frac {R}{R_p} \right )^3
                                 \cos \left ( \phi-\phi_p \right ).
\end{eqnarray}
This can be expressed in the angle-action variables as (cf.\ eq.\ \ref{eq:V1-AA-space})
\begin{eqnarray}
V_{\rm e} \simeq {\rm Re} \sum_{l=-\infty}^{+\infty} Q(t) 
\tilde h_{{\rm e},l}(\Jvec) e^{{\rm i}(l w_R+w_{\phi} )},
\label{eq:Fourier-expansion-Ve-m1}
\end{eqnarray}
where 
\begin{eqnarray}
Q(t) = -\frac{3}{8} \left ( \frac{M_p}{M_\star} \right ) \frac{1}{[R_p(t)]^4} e^{-{\rm i}\phi_p(t)},
\end{eqnarray}
is the time-varying part of the external perturbation, and 
\begin{eqnarray}
\tilde h_{{\rm e},l}(\Jvec)=\frac{1}{2\pi} \oint R^3
\cos \left [ l w_R + (w_{\phi} - \phi ) \right ] ~\dif w_R.
\end{eqnarray}
For a parabolic encounter with minimum distance $R_{p,{\rm min}}$ and gravity parameter 
$\bar M=1+M_p/M_{\star}$, the true anomaly $\phi_p$ and radial distance $R_p$ are 
computed through the following equations:
\begin{eqnarray}
t(\phi_p) \!\!\! &=& \!\!\! \frac{\sqrt{2}}{\omega_p} \left [ \tan \left ( \frac{\phi_p}{2} \right )+
                   \frac 13 \tan^3 \left ( \frac{\phi_p}{2} \right ) \right ], 
                   \label{eq:time-vs-true-anomaly} \\
R_p \!\!\! &=& \!\!\! \frac{2R_{p,{\rm min}}}{1+\cos(\phi_p)},~~ \bar M= \omega_p^2 R_{p,{\rm min}}^3.
\label{eq:parabola-equation}
\end{eqnarray}

The effect of the external perturbation on the evolution of $f_1$ inside the $n$th 
element is determined by the Galerkin projection of the Poisson
bracket $-[f_0,V_{\rm e}]$ as 
\begin{eqnarray}
4\pi^2 {\rm i} \Zvec^{n'}_{l'}(t) \!\!\! &=& \!\!\! -\int \!\! \int
\Evec_{l'}^{\rm T}(n',\Jvec) [f_0,V_{\rm e}]  \nonumber \\ 
\!\!\! & {} & \!\!\! \qquad \times
e^{-{\rm i}(l'w_R+w_{\phi})} 
~\dif^2 \Jvec ~\dif^2 \wvec.
\label{eq:projected-external-potential-1}
\end{eqnarray}
Substituting from (\ref{eq:Fourier-expansion-Ve-m1}) into 
(\ref{eq:projected-external-potential-1}) and performing 
the integral over the angle space gives the 2-vector
\begin{eqnarray}
\Zvec^n_l(t) = Q(t) \int
\left ( l \frac{\partial f_0}{\partial J_R} + 
\frac{\partial f_0}{\partial J_{\phi}} \right )
\tilde h_{{\rm e},l}(\Jvec) \Evec_{l}^{\rm T}(n,\Jvec) ~\dif^2 \Jvec,
\label{eq:projected-external-potential-2}
\end{eqnarray}
whose components ($Z^n_{1,l}$ and $Z^n_{2,l}$) are, respectively, 
the contribution of the disturbing force to the inner and outer 
nodes of the $n$th ring element in the configuration space. 
The disturbance at the $j$th ring node thus reads  
\begin{eqnarray}
Z_{j,l}(t) = \left \{
\begin{array}{ll}
Z^j_{1,l} ~, & j=1, \\ \\
Z^j_{1,l}+Z^{j-1}_{2,l} ~, & 1<j<N+1, \\ \\
Z^N_{2,l} ~, & j=N+1,
\end{array}
\right.
\end{eqnarray}
and we obtain
\begin{eqnarray}
\Zvec_l(t)= \left [
\begin{array}{lllll}
Z_{1,l} & Z_{2,l} & \ldots & Z_{N,l} & Z_{(N+1),l}
\end{array}
\right ]^{\rm T}.
\end{eqnarray}
Defining $\Fvec_l = \Umat_1^{-1}(l) \cdot \Zvec_l$, the global forcing vector is assembled as
\begin{eqnarray}
\Fvec(t) \!\!\! &=& \!\!\!
\left [
\begin{array}{lllll}
\ldots & \Fvec^{\rm T}_{-1}(t) & \Fvec^{\rm T}_0(t) & \Fvec^{\rm T}_{+1}(t) & \ldots 
\end{array}
\right ]^{\rm T}, \nonumber \\
 \!\!\!\ &=& \!\!\! Q(t) \gvec,
\end{eqnarray}
where $\gvec$ is a constant vector. For each Fourier number $l$, 
we have $N_{\rm t}$ unknown DFs collected in the vector $\zvec_{\lvec}(t)$. 
The Fourier series in terms of $w_R$ is usually truncated at some 
$l_{\rm min} <0$ and $l_{\rm max} >0$. We thus have 
${\cal N}=(l_{\rm max}-l_{\rm min}+1)\times N_{\rm t}$ unknown 
DFs that we collect in the ${\cal N}$-vector $\zvec(t)$. Similarly, 
$\Fvec(t)$ and $\gvec$ are ${\cal N}$-dimensional vectors. 

Any excited wave is a superposition of all eigenmodes of 
(\ref{eq:eigensystem}):
\begin{eqnarray}
\zvec(t)=\sum_{j=1}^{{\cal N}} q_j(t) \zvec^{(j)},
\label{eq:expansion-modal}
\end{eqnarray}
but not all eigenvectors $\zvec^{(j)}$ are physical. Increasing the number of elements 
increases the accuracy of the eigenvalues and eigenvectors describing the isolated 
oscillatory modes but also adds spurious and/or singular modes. Such non-physical 
modes can contribute noise to the calculated disc response. To keep only physical 
modes, we introduce
\begin{eqnarray}
\qvec \!\!\! &=& \!\!\! \left [
\begin{array}{llll}
q_1 & q_2 & \ldots & q_{{\cal N}}
\end{array}
\right ]^{\rm T}, \nonumber \\
\Mmat \!\!\! &=& \!\!\! \left [
\begin{array}{llll}
\zvec^{(1)} & \zvec^{(2)} & \ldots & \zvec^{({\cal N})}
\end{array}
\right ], \nonumber 
\end{eqnarray}
and express (\ref{eq:expansion-modal}) in the matricial form 
$\zvec=\Mmat \cdot \qvec$. This is substituted into (\ref{eq:forced-linear-equations})
to obtain 
\begin{eqnarray}
\frac{\dif}{\dif t}\qvec = -{\rm i} \Jmat
\cdot \qvec +{\rm i} Q(t) \Mmat^{-1} \cdot \gvec, 
\label{eq:decoupled-linear-ODEs}
\end{eqnarray}
where $\Jmat = \Mmat^{-1} \cdot \Amat \cdot \Mmat$ is a diagonal matrix---or a Jordan 
form if there are degenerate eigenvalues \citep{P01}---whose elements are the 
eigenfrequencies of (\ref{eq:eigensystem}). The diagonalizing matrix $\Mmat$ is often 
called the modal matrix.

\begin{figure}
\centerline{\hbox{\includegraphics[width=0.45\textwidth]{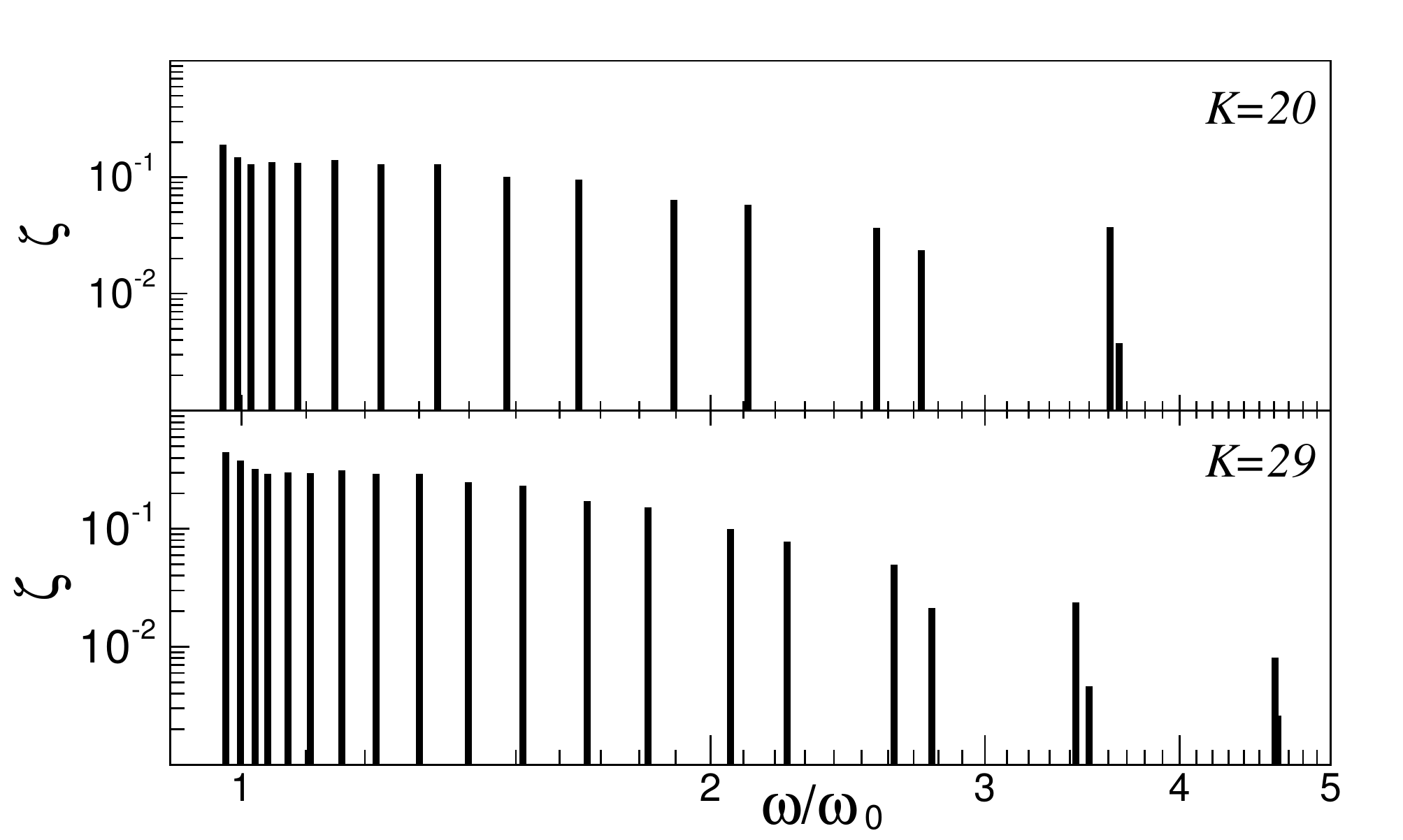}} }
\caption{The forcing vector components $\zeta_j$ for the modes of the
  discs having $\mu=0.05$ and $K=20$ and $29$. 
The corresponding frequency spectra have been plotted in 
Figure \ref{fig4}.}
\label{fig8}
\end{figure}

We call the ${\cal N}$-dimensional vector $\bfzeta=\Mmat^{-1} \cdot \gvec$ the 
forcing vector and rewrite (\ref{eq:decoupled-linear-ODEs}) in terms of its components:
\begin{eqnarray}
\frac{\dif}{\dif t} q_j(t) = -{\rm i} \omega_j q_j(t) +{\rm i} Q(t) 
\zeta_j,~~ j=1,2,\ldots,{\cal N}.
\label{eq:decoupled-linear-ODEs-components}
\end{eqnarray}
Equations associated with non-physical $\omega_j$ can now be dropped
from (\ref{eq:decoupled-linear-ODEs-components}) and we find both 
the homogeneous and particular solutions,
\begin{eqnarray}
q_j(t) = e^{-{\rm i}\omega_j (t-t_0)} q_j(t_0) +
{\rm i} \zeta_j \int_{t_0}^{t} Q(\tau) e^{-{\rm i}\omega_j (t-\tau)} 
~ \dif \tau,
\end{eqnarray}
for $j=1,2,\ldots,{\cal N}_p$ with ${\cal N}_p$ being the number of physical modes. 
Fly-by perturbations begin at $t=t_0=-\infty$ ($\phi_p=-\pi$) with $q_j(t_0)=0$ ($\forall j$). 
Consequently, using the orbit equations (\ref{eq:time-vs-true-anomaly}) and 
(\ref{eq:parabola-equation}), and defining $\beta=\omega_p/\omega_j$, 
we arrive at 
\begin{eqnarray}
q_j(t,\phi_p) \!\!\! &=&  \!\!\! - {\rm i} \frac{3\sqrt{2}(\bar M-1) }{64 \bar M^{4/3}}\omega_j^{5/3} 
\zeta_j Q_j e^{-{\rm i}\omega_j t}, \label{eq:solution-qj-vs-phi_p} \\
Q_j \left ( \phi_p,\beta \right ) \!\!\! &=& \!\!\!  
\beta ^{5/3}  \int_{-\pi}^{\phi_p} \left(1+\cos\xi\right)^2  e^{ {\rm i} \omega_j t(\xi)- {\rm i} \xi} ~ \dif \xi,
\end{eqnarray}
which leaves behind the permanent oscillation 
\begin{eqnarray}
q_j(t) =  - {\rm i} \frac{3\sqrt{2}(\bar M-1)}{64 \bar M^{4/3}} \omega_j^{5/3}
\zeta_j Q_j(\pi,\beta) e^{-{\rm i}\omega_j t},  \label{eq:solution-qj-infinity}
\label{eq:response-qj}
\end{eqnarray}
when the encounter ends at $\phi_p=+\pi$. The integrands of the real and imaginary parts of 
$Q_j(\pi,\beta)$ are, respectively, even and odd functions of $\phi_p$ over the interval 
$[-\pi,+\pi]$. One thus obtains ${\rm Im}[Q_j(\pi,\beta)]=0$. The real part of 
$Q_j(\pi,\beta)/\beta^{5/3}$ is positive-definite, and therefore, 
a necessary and sufficient condition for the excitation of the $j$th oscillatory mode is that 
the corresponding component of the forcing vector $\zeta_j\not =0$. The asymptotic forms are 
\begin{eqnarray}
Q_j(\pi,\beta) \approx 2\pi \beta^{5/3},
\label{eq:betabig}
\end{eqnarray}
for $\beta \gg 1$ and 
\begin{eqnarray}
Q_j(\pi,\beta) \approx \frac{2^{15/4} \pi^{1/2}}{3} \beta^{1/6} \exp \left ( -\frac{2\sqrt{2} }{3\beta}  \right ),
\label{eq:betasmall}
\end{eqnarray}
for $\beta \ll 1$. The exponential decay for small $\omega_p$ is due to adiabatic
invariance. 

We have computed $\bfzeta$ for all parent modes of Figure \ref{fig4}, and have plotted its components 
$\zeta_j$ in Figure \ref{fig8} for two $\mu=0.05$ models with different mean eccentricities. The results 
for child modes and other models are similar. In our models $\zeta_j$ is larger for modes with low 
frequencies. This can be understood as a competition between two effects seen in Figure \ref{fig5}: 
(i) as the mode frequency decreases, the number of its nodes increases, so the coupling of the 
mode to a smooth external field is reduced; (ii) as the frequency decreases, 
the outermost peak of $X(R)$ shifts to a larger radius and hence contributes more to the term 
$-[f_0,V_{\rm e}]$ in (\ref{eq:linearized-Vlasov})---recall that $V_{\rm e}\sim R^3$. In general the 
second effect wins, causing the coupling, as measured by $\bfzeta$, to be larger for low-frequency
modes. 

We also find that the range of $\bfzeta$ is similar for all $K$ models
(Fig.\ \ref{fig8}). This shows that the response 
of a near-Keplerian disc is not sensitive to its mean eccentricity:
$m=1$ slow modes in warm and cold discs have 
an equal chance of being excited by encounters.  

\begin{figure}
\centerline{\hbox{\includegraphics[width=0.47\textwidth]{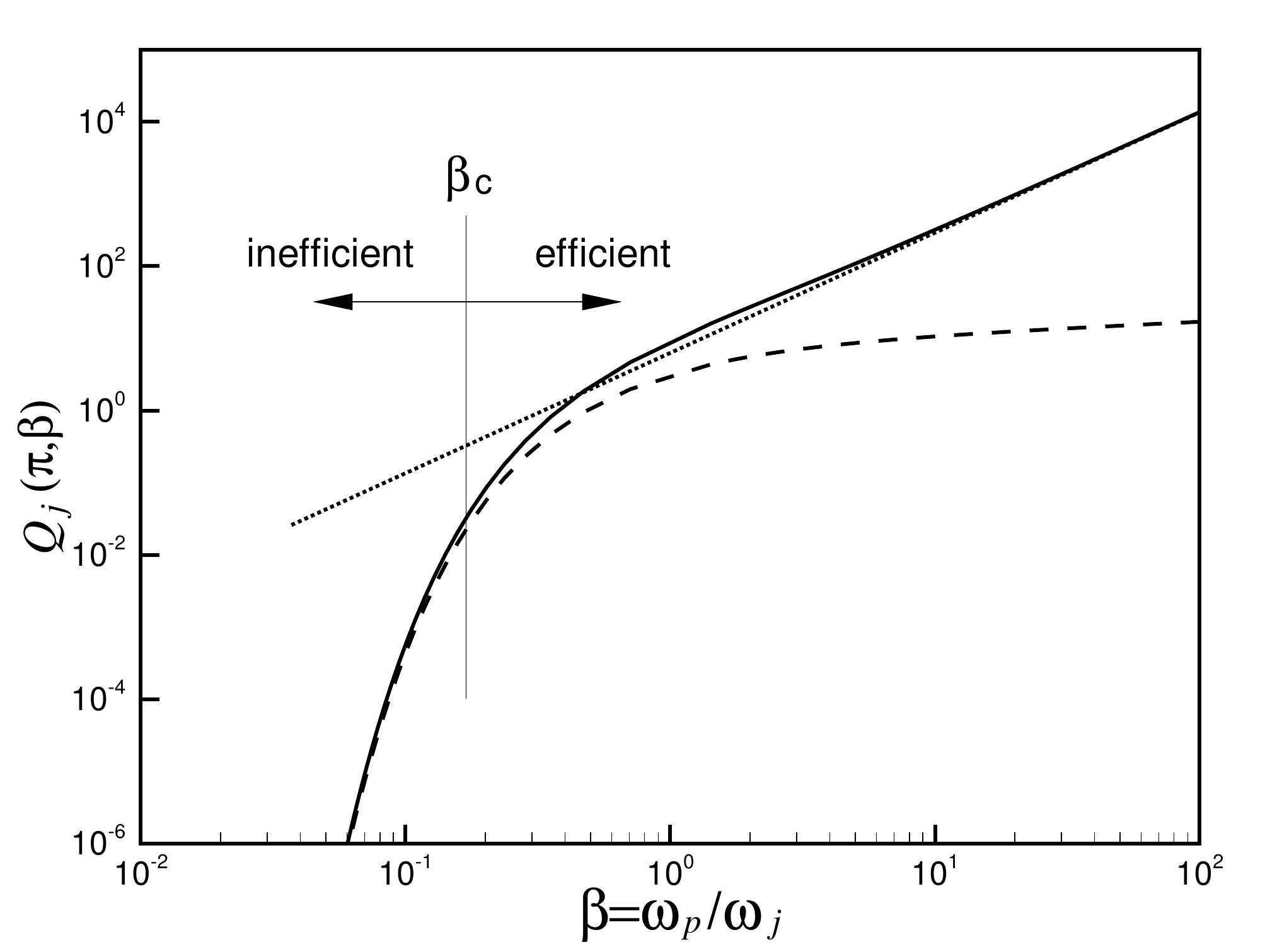}} }
\caption{
The profile of $Q_j(\pi,\beta)$ (solid line) together with its 
analytic asymptotes when $\beta \rightarrow 0$ (dashed line; eq.\
\ref{eq:betasmall}) and $\beta \rightarrow \infty$ (dotted line; eq.\ \ref{eq:betabig}). 
The excitation is efficient for $\beta > \beta_c$ with
$\beta_c=0.169367$ (see text for definition).}
\label{fig9}
\end{figure}

The excitation efficiency of modes is determined by the function $Q_j(\pi,\beta)$, 
which has been plotted in Figure \ref{fig9}. The excitation of mode $j$ is inefficient 
for $\omega_p \lesssim \beta_c \omega_j$ where we have defined the critical frequency ratio
$\beta_{c}=0.169367$ as the point where $Q_j(\pi,\beta)$ drops to 10\% of its value 
predicted by its $\beta \rightarrow \infty$ asymptote (eq.\ \ref{eq:betabig}). For a given mass parameter $\bar M$, 
a perturber can only excite mode $j$ efficiently if its orbit has periastron 
$R_{p,{\rm min}}\lesssim [ \bar M/ (\beta_c^2 \omega_j^2) ]^{1/3}$. Faster modes 
have larger $\omega_j$, and therefore need closer encounters to be excited. 

\begin{figure*}
\centerline{ \hbox{\includegraphics[width=0.5\textwidth]{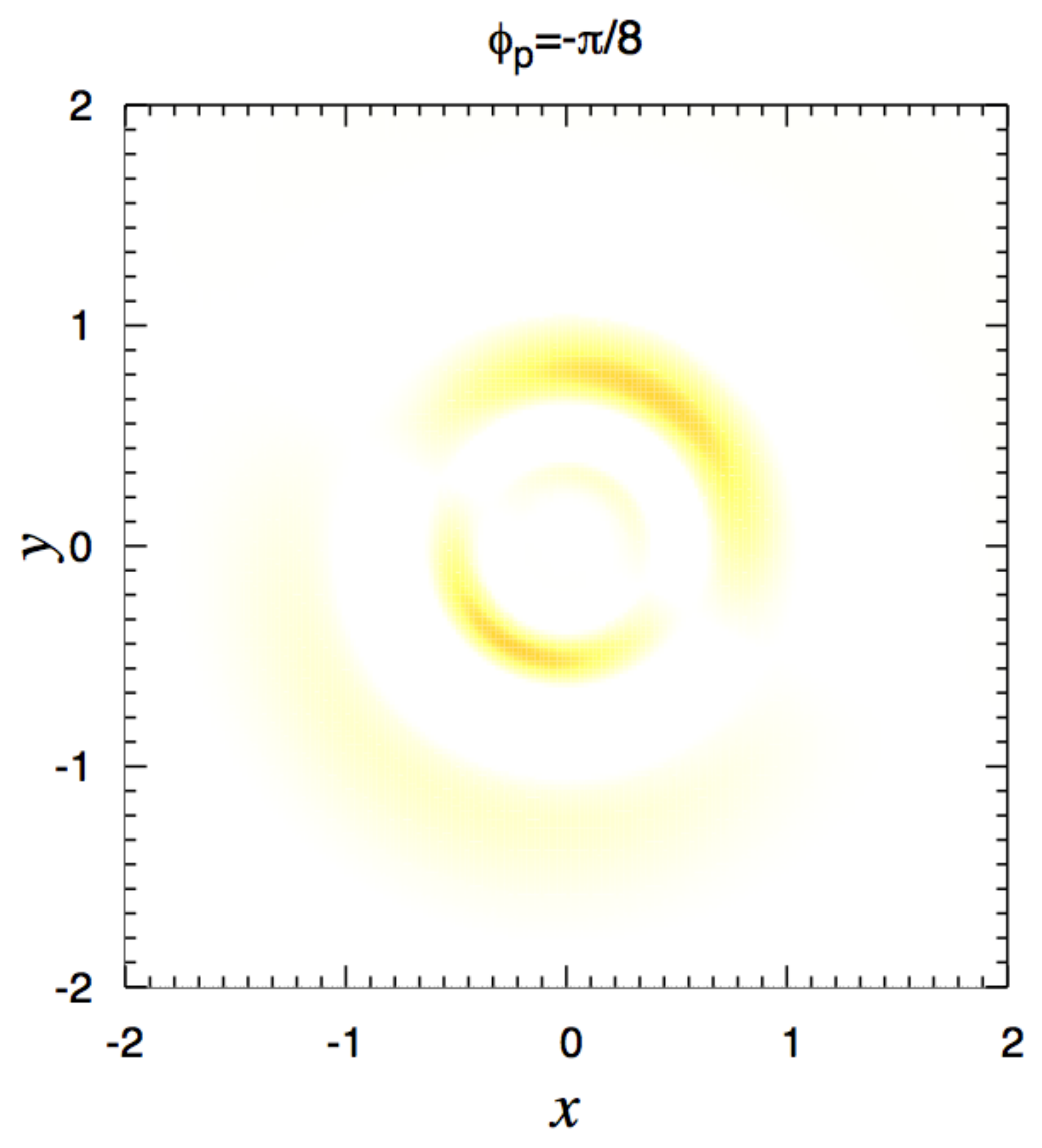}}
	           \hbox{\includegraphics[width=0.5\textwidth]{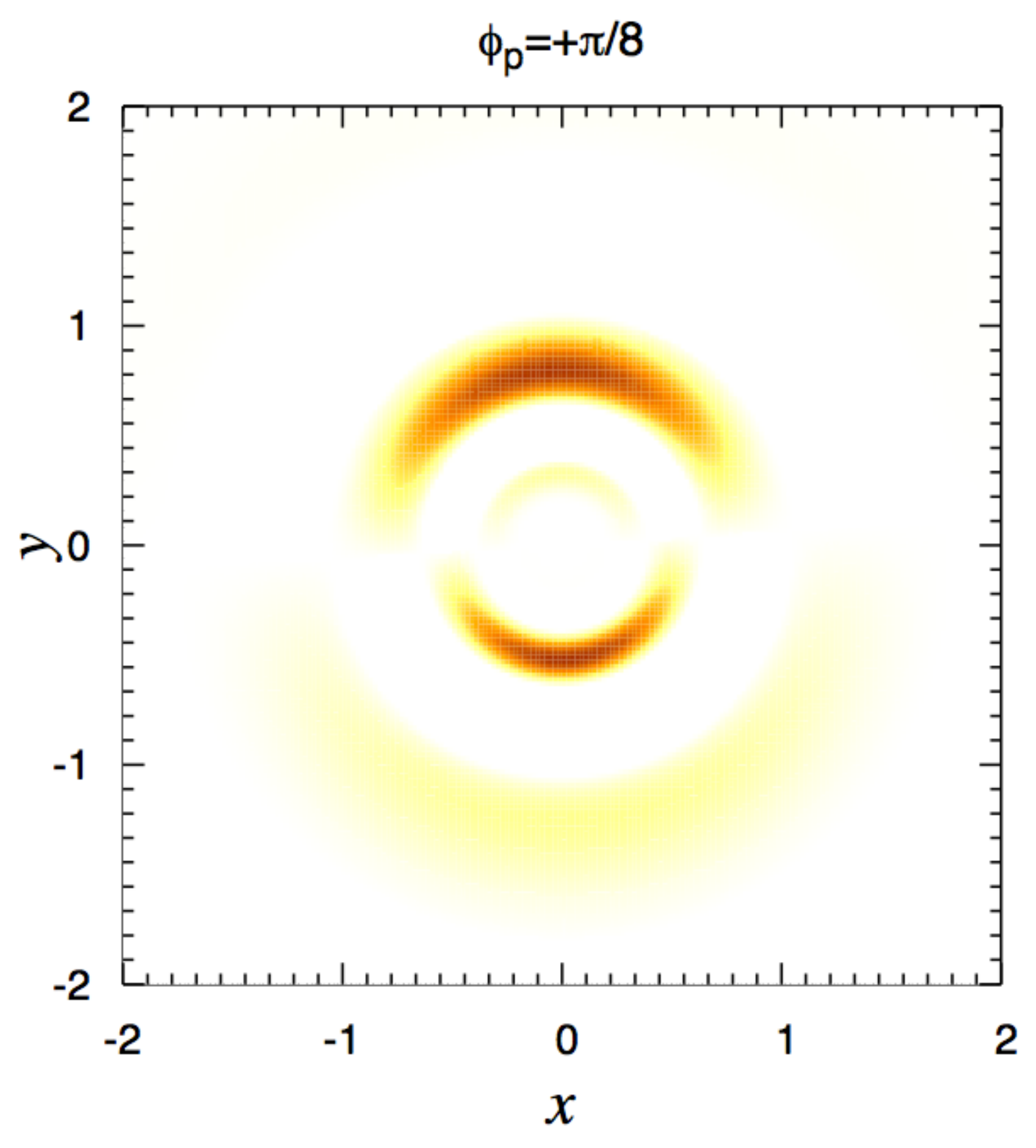}}}
\centerline{ \hbox{\includegraphics[width=0.5\textwidth]{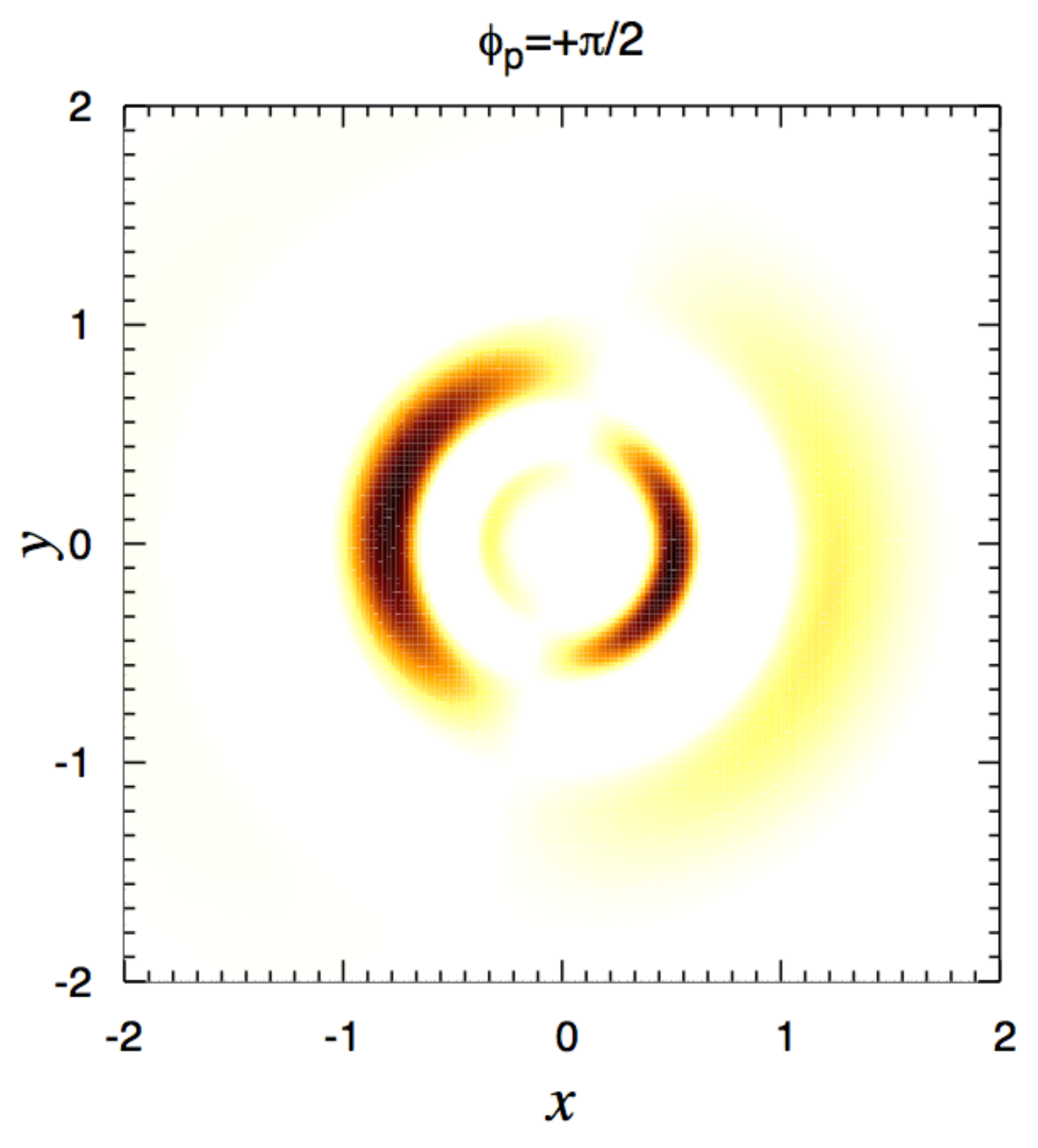}}            
                    \hbox{\includegraphics[width=0.5\textwidth]{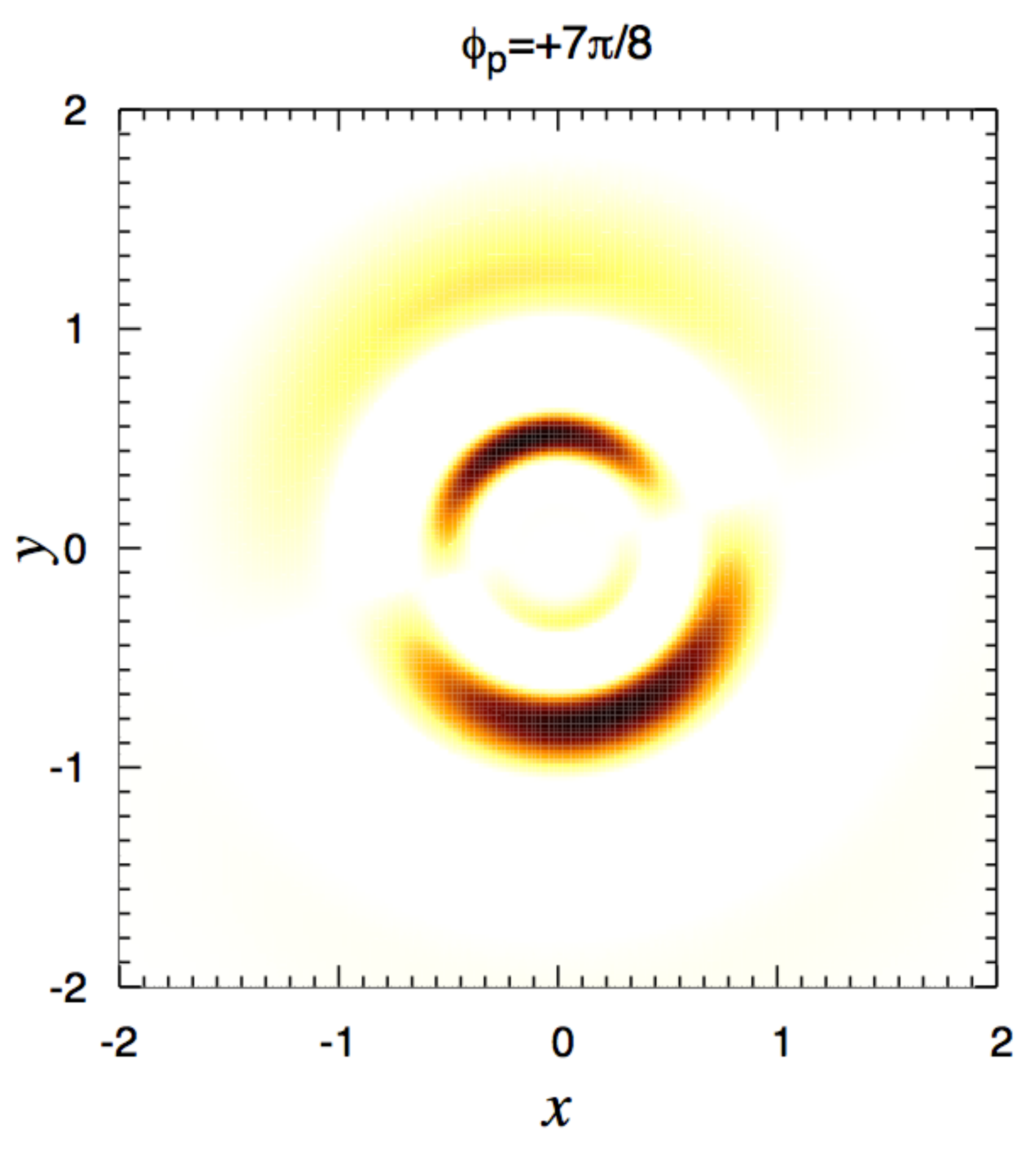}}
	          }

\caption{The evolution of mode ${\rm D}_1$ in a disc with mean
  eccentricity $\bar e=0.159$ as a perturber on a parabolic 
orbit encounters the star-disc system. $\phi_p$ is the azimuthal angle of the flying-by star; 
the line $\phi_p=0$ coincides with the $x$-axis and with the direction
of periastron. We have $\phi_p=(-\pi,+\pi)$ for $t=(-\infty,+\infty)$.
The contours show the positive part of the response density. The periastron distance of the 
perturber is $R_{p,\rm min}=188.58(1+M_p/M_\star)^{1/3}(M_{\rm d}/10^{-3}M_\star)^{-2/3}$.}
\label{fig10}
\end{figure*}

In Figure \ref{fig10}, we illustrate the excitation of mode ${\rm D}_1$ 
at four azimuths during the fly-by of a disc with mass ratio $\mu=0.025$, for an encounter with the parameter $\omega_p=\sqrt{2} \omega_{{\rm D}_1}=0.0096538$. 
The interaction begins at $\phi_p=-\pi$ ($t=-\infty$) and ends at $\phi_p=+\pi$ ($t=+\infty$). 
We plot the positive part of the response density:
\begin{eqnarray}
\Sigma_{1,{\rm D}_1}(R,\phi,t) \!\!\! &=& \!\!\! {\rm Re} \Big \{ q_{j}[t,\phi_p(t)]~ e^{{\rm i}m\phi-{\rm i} \omega_{j} t } \nonumber \\
\!\!\! &{}& \!\!\! \quad \times \sum_{n=1}^{N} H_n(R) \Gvec_n(R) \cdot \bvec^{(j)}_n \Big \},
\end{eqnarray}
where $j$ corresponds to mode ${\rm D}_1$ and the vectors $\bvec^{(j)}_n$ are extracted from 
$\dvec^{(j)}=\sum_l \Fmat(l)\cdot \zvec^{(j)}_l$ as we did in equation (\ref{eq:find-b-from-d}). 

We remark that stable modes always rotate with a constant angular velocity, but the 
perturber-star centreline has a variable angular velocity. Therefore the perturber may lead 
or lag the maximum response in azimuth, and the maximum response may occur some
time after closest approach. 

In general, of course, the close passage of a perturber will excite multiple modes. 
The main visual difference between a single-mode and multi-mode response is the occurrence 
of long-period beating patterns in the latter case. We have constructed animations of the 
evolution of the multi-mode pattern during an encounter and the beating can be quite
striking 
({\texttt http://www.youtube.com/watch?v=ZTyXK7H6Q8E})
This animation is for a model with $K=10$ and $\mu=0.025$, and the 11 modes with the 
highest frequencies are participating in the response.

\section{Application to debris discs and galactic nuclei}

\label{sec:appl}

We have shown that low-mass, near-Keplerian, collisionless discs can
support stable, long-lived, large-scale slow modes. The most
prominent of these are expected to have azimuthal wavenumbers
$m=1$ and $m=2$. 

\subsection{Debris discs}

The existence of slow modes implies that debris discs can support
waves in the planetesimal population that provides most of the
disc mass, and suggests that collisions in this non-axisymmetric
distribution could generate non-axisymmetric dust distributions that
would be visible in thermal emission or scattered light.

Non-axisymmetric structures in debris discs are normally assumed to be
produced by planets, but our results imply that some or even most of these
structures may be density waves. Specific examples include:

\begin{itemize}

\item $\beta$ Pictoris: Scattered starlight reveals that this star is
  surrounded by a debris disc extending to $\gtrsim 1000$ AU. The disc
  is brighter on one side than the other, perhaps due to an $m=1$ slow
  mode, and also contains brightness enhancements that could be due to
  shorter-wavelength density waves. The disc exhibits warps or tilted
  rings at various radii \citep{heap00,wah03}; although the present
  paper examines only in-plane slow modes, there should also be slow
  bending modes, and these provide a possible explanation for the
  warps. There is a $10 M_J$ planet orbiting at $\sim 10$ AU in the
  $\beta$ Pic system \citep{lag10} but it is far from clear that this
  is the cause of the warps and other features; several authors
  have argued that the asymmetries provide evidence for two or even
  three planets \citep{frei07,currie} but it is implausible to invoke
  a new planet for every feature.

\item Fomalhaut: This star is surrounded by a ring of dust with a
  sharp inner edge at 130 AU. The centre of the ring is offset by 15
  AU from the host star, implying an eccentricity of 0.11; the ring is
  narrowest at apastron, implying that the eccentricity declines with
  radius \citep{KGC05}. \cite{qui06} stressed that these features
  could be produced by a planet orbiting just inside the ring, and
  a possible planet was subsequently discovered \citep{K08}. The
  eccentricity of the ring could be forced by the planet or a slow
  density wave, depending on whether the planet mass or
  ring mass is larger. The sharp inner edge of the ring is most likely
  due to the planet.

\item Vega: Observations at a variety of wavelengths between $350\mum$
  and 1.3 mm reveal a face-on dust ring of radius $\sim 100$ AU,
  dominated by two clumps (see \citealt{marsh} for a summary of the
  data, and beware that \citealt{pietu} question the reality of
  non-axisymmetric structure in the disc). The clumps are usually
  ascribed to dust trapped in a resonance with an unseen planet \citep[e.g.,][]{kuchner,wy03}, but
  $m=1$ and $m=2$ slow density waves provide an alternative
  explanation. Within a few years we may be able to distinguish these
  hypotheses by measurements of the motion of these clumps relative to
  the host star: the expected angular speed of the planet is $\sim
  1^\circ\,\mbox{yr}^{-1}$ while slow modes should have negligible
  pattern speeds. 

\item $\epsilon$ Eridani: A nearly face-on ring of dust surrounds this
  star at $\sim 60$ AU. The disc exhibits several clumps and a
  lopsided brightness distribution in images at $450\mum$ and
  $850\mum$ \citep{gre05}. Some but not all of these peaks may be
  background sources. Models in which the clumps are due to resonances
  with a planet are described by \cite{ozernoy}, \cite{qt02}, and
  \cite{deller}. These features could be due to slow modes, but the
  presence of density maxima at several azimuths would require that
  more than one mode was present. Resonance models predict angular
  velocities around the host star of about $1^\circ\,\mbox{yr}^{-1}$.

\item HR 4796A: There is an edge-on debris ring
  $\sim 80$ AU from the host star. One ansa of the ring is brighter,
  hotter, and at smaller radius than the other
  \citep{tel00,moe11}. This asymmetry is most naturally explained by
  an eccentric dust ring ($e\simeq0.06$); at periastron the dust is
  closer to the star and therefore hotter and brighter (``pericentre
  glow'', \citealt{wy99}). The ring eccentricity is usually assumed to
  be excited by secular perturbations from a nearby planet but an
  $m=1$ slow mode of the disc is an alternative. The mode
  might be excited by the companion star HR 4796B, currently at a
  projected separation of $\sim 500$ AU.

\item AB Aurigae: Near-infrared images reveal a debris disc of over
  1000 AU radius. The
  disc shows spiral arms at radii of several hundred AU, some of which
  are also seen at submm wavelengths, as well as rings, gaps, and
  clumps at smaller radii \citep{hash11}. As in the case of other systems, the
  features could be due to planets or slow modes, and these hypotheses
  can be distinguished by proper-motion measurements. 

\item $\eta$ Corvi: The disc surrounding this star appears at
  $450\,\mum$ as two equally bright peaks equidistant from the host
  star at 100 AU; these can be modelled either as the ansae of an
  edge-on axisymmetric ring or as a more face-on disc containing dust
  trapped in a resonance with a planet \citep{wy05}. A third
  possibility is an $m=2$ slow mode in a face-on disc. 

\item HD 141569A: \cite{cla03} have detected strong spiral structure
  in the debris disc around this star, at about 400 AU radius. They
  suggest that the spiral may be excited by tides from nearby
  stars. There is also a gap in the disc at about 250 AU radius; both
  of these features might be due to planets but only if planets can
  form at radii exceeding 200 AU. \cite{wyatt05} has suggested that
  the spiral could be caused by a Jupiter-mass planet on an eccentric
  orbit ($e\simeq0.2$, $a\simeq 250$ AU) but slow modes provide a more
  economical explanation, especially given the difficulties of
  forming planets at such large distances. 

\item HD 100546: This disc exhibits an apparent dark hole and bright
  clump at about 30 AU from the host star \citep{quanz}. These features
  could be due to an orbiting planet or a slow density wave. The
  Keplerian motion at this radius is about
  $3^\circ\,\mbox{yr}^{-1}$. At much larger radii, $\sim 250$ AU, the
  disc exhibits spiral structure \citep{grady}. Possible explanations
  include a planet at several hundred AU from the star or density
  waves excited by a passing star. The latter possibility was discussed
  by \cite{quillen05} but dismissed because their estimated lifetime
  for the spiral structure was only $\sim 10^4$ yr and no suitable
  nearby star could be found; the results of the present paper imply
  that the structure could last for a much longer time---perhaps as
  long as the 10 Myr age of the star---so the chance of a
  suitable encounter in the past is much larger.

\item HD 61005: This star is surrounded by an asymmetric edge-on
  debris disc of radius $\sim 60$ AU. The asymmetry can be modelled as
  a mean eccentricity of 0.05, but there are no planets more massive
  than $\sim 3$ Jupiter masses close to the ring \citep{buenzli}.

\item HD 15115: This star hosts an edge-on debris disc; the dominant
  thermal emission from the disc arises at radii $\sim 35$ AU but the
  disc is visible to much larger radii. The surface brightness of the
  east side of the disc is about 1 mag fainter than the west side at a
  given radius and the surface-brightness distribution perpendicular
  to the disc midplane is asymmetric on the west side \citep{kalas07};
  both features can arise naturally from an $m=1$ distortion.

\item HD 107146: There is a dust ring at 100 AU that exhibits clumps
  and a lopsided brightness distribution in 1.3 mm images
  \citep{cor09}. These might be due either to a planetary resonance or
  to slow density waves; however, $880\mum$ observations with similar
  resolution do not confirm the existence of the clumps
  \citep{hughes11}.

\end{itemize}

\subsection{Discs in galactic nuclei}

The results of this paper also illuminate our understanding of stellar
discs in galactic nuclei. They can be applied directly to such discs
if the apsidal precession is dominated by the self-gravity of the
disc, rather than relativistic effects or the gravitational field from
a spherical stellar population in the nucleus. 

The apparent `double' nucleus of M31 is most likely a
stellar disc that has been distorted by a large-amplitude slow mode
(see \citealt{pt03}, \citealt{ss04}, and references therein). Such
modes arise naturally in N-body simulations \citep{sellwood}. They can
be excited by gas inflow and star formation in the central few parsecs
of the galaxy \citep{hq10} or by instabilities induced by a small
population of counter-rotating stars \citep{touma}. Slow modes may
also play a central role in feeding supermassive black holes
\citep{hq10a}.

\section{Discussion}
\label{sec:disc}

The finite element formulation has enabled us to explore the modal
spectrum of low-mass near-Keplerian collisionless discs, and to
calculate the corresponding mode shapes for a wide range of initial
radial dispersions (rms eccentricities). Our method also yields
moments of the distribution function, which provide the evolutionary
equations for energy and angular-momentum transport in perturbed
discs, and allows the accurate representation of modes that contain a
singular resonant component (e.g., Fig.\ \ref{fig6}). 

We find that near-Keplerian discs support `slow' modes, that is, modes
for which the eigenfrequency or pattern speed is proportional to the
disc mass.  WKB analysis shows that these modes are
closely related to the $p$-modes found by \cite{T01} in cold
near-Keplerian discs with softened gravity. Both our numerical results
and analytic arguments imply that there are no unstable slow
modes. All slow modes in the discs we have examined are prograde
(positive pattern speed). Slow modes can exist with arbitrary
azimuthal wavenumber $m$, but modes with $m=1$ and $m=2$ have the
largest scale and are the easiest to excite by an external perturber. 

The eigenmodes of the linearized CBE bifurcate from the degenerate
leading/trailing modes predicted by WKB theory. The modes are
degenerate for cold discs and split into close (in frequency) pairs as the mean
eccentricity grows, until for $\bar e \gtrsim 0.2$ there is no
apparent pairing in frequency space (see Fig.\ \ref{fig4}). 

Some of the non-axisymmetric structure that is commonly observed in
debris discs, such as clumps, lopsided rings, and spiral arms may be
due to slow modes, perhaps excited by the fly-by of a passing star or
binary companion. These features are normally ascribed to hypothetical
massive planets embedded in the disc. The two hypotheses can be
distinguished in some discs by monitoring the motion of these features
over decade timescales: many features associated with planets should orbit
the host star at a pattern speed that is not far from the Keplerian
angular speed of the planet, whereas slow modes should have negligible
pattern speeds. Structures induced by modes and
planets may also be distinguishable in the future by high-resolution
far-infrared observations by interferometers or large single-dish
telescopes (e.g., ALMA or CCAT). 

Future theoretical work should include the exploration of slow bending
modes and of the behavior of slow modes in thick discs that resemble
the discs seen in galactic nuclei. 

\section*{Acknowledgements}

MAJ thanks the School of Natural Sciences at the Institute for Advanced Study, 
Princeton, for their generous support.  This research was supported in part by 
NSF grant AST-0807432 and by NASA grants NNX08AH83G and NNX11AF29G. 
We thank the anonymous referee for a useful report that helped to improve our 
presentation and understanding. 


\appendix 

\section{WKB analysis}

The dispersion relation for a collisionless disc can be computed
analytically in the WKB or short-wavelength approximation \citep{BT08}
\begin{align}
1=&\frac{2\pi G\Sigma_0|k|}{\Omega_R^2}{\mathcal
  G}\left[\frac{\omega-m\Omega_{\phi} }{\Omega_R},\left ( \frac{\sigma_Rk}{\Omega_R}\right )^2\right],
\notag \\
\mbox{where}&\quad
{\mathcal
  G}(s,\chi)=\frac{2}{\chi}e^{-\chi}\sum_{n=1}^\infty\frac{I_n(\chi)}{1-s^2/n^2}.
\label{eq:lsk}
\end{align}
Here $\Omega_R$ and $\Omega_\phi$ are the orbital frequencies (eq.\
\ref{eq:orbfreq}), $I_n(\chi)$ is a modified Bessel function, $m$ is the azimuthal wavenumber, $\sigma_R$ is the
radial velocity dispersion, and the perturbed surface density is
assumed to vary as $\Sigma_1(R,\phi,t)\propto
\exp[i(m\phi+\int^rk(r')dr'-\omega t)]$. The dispersion relation
(\ref{eq:lsk}) is valid if $\sigma_R\ll\Omega_\phi R$ and $|k|R\gg
1$. 

\begin{figure*}
\centerline{\hbox{\includegraphics[width=0.45\textwidth]{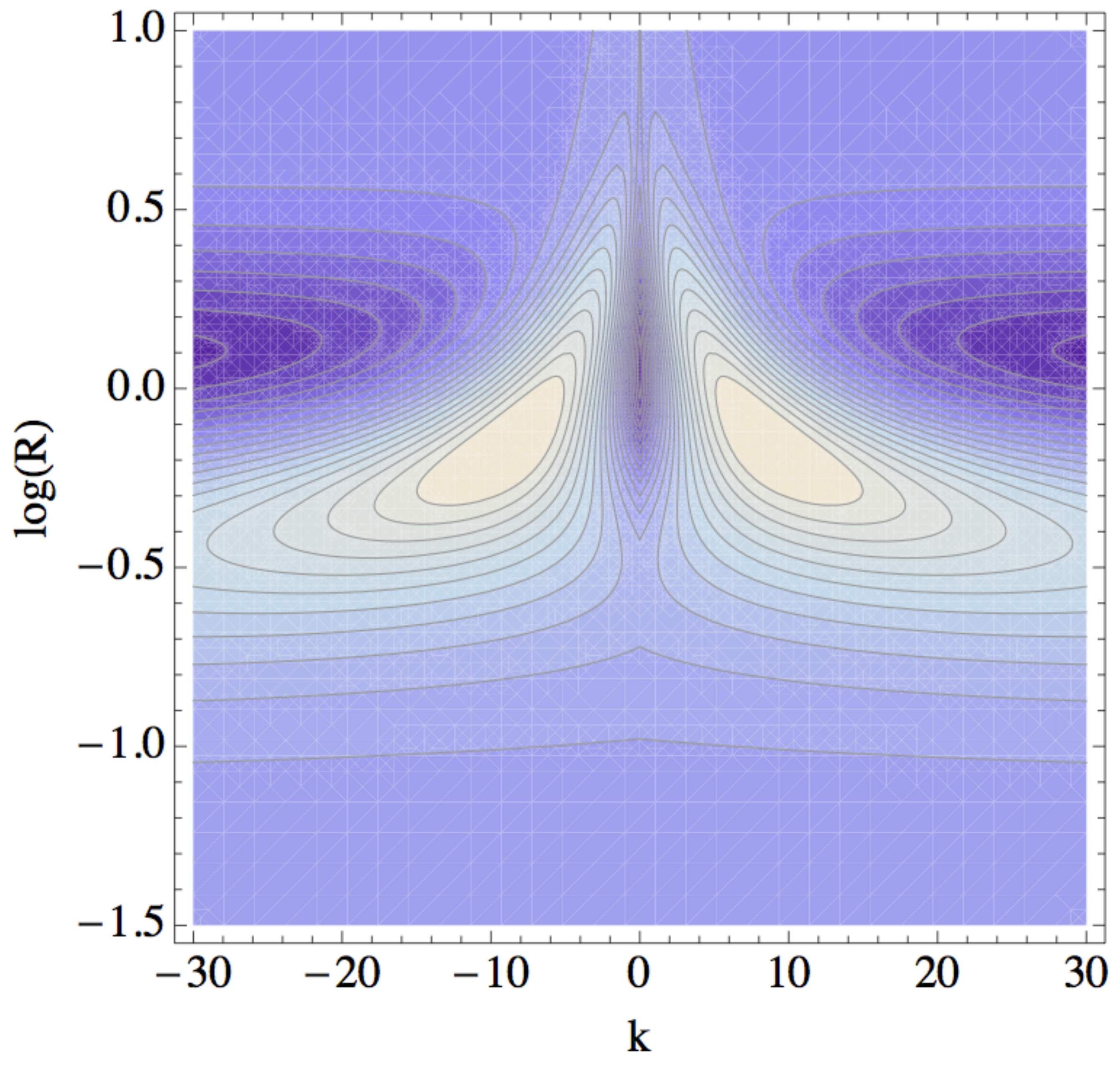}}
\hspace{0.1in}
                  \hbox{\includegraphics[width=0.45\textwidth]{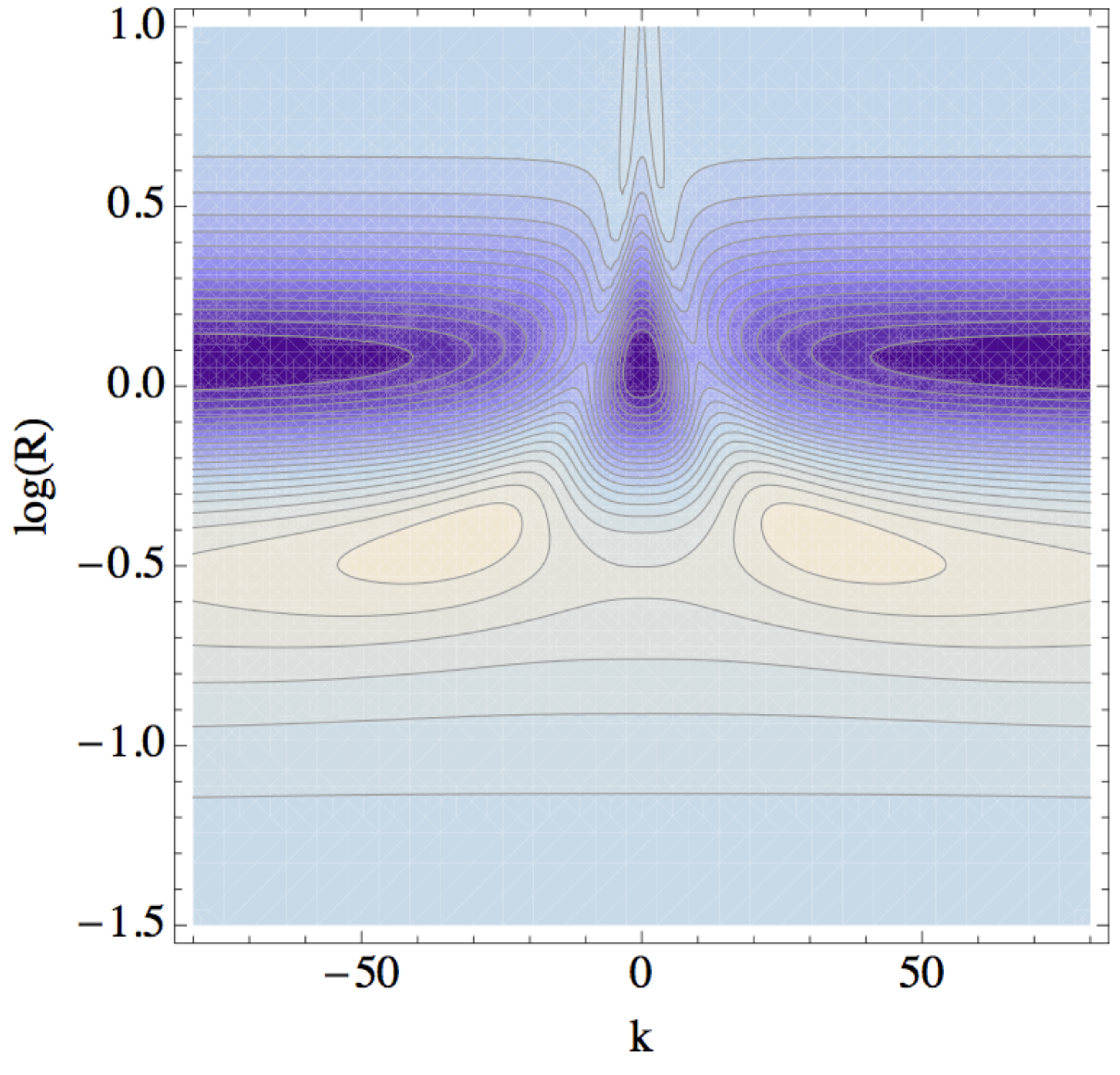}}
	    }
\caption{Contours of constant $\omega$ in the WKB approximation, from
  equation (\ref{eq:lskslow}) for $m=1$ (left) and $m=2$ (right). The disc has mean eccentricity $\bar e=0.159$. 
  The horizontal axis is the wavenumber and the vertical axis is the
  log of the radius (base 10).   Lighter shades correspond to higher values of $\omega$. }
\label{fig:cont}
\end{figure*}

For low-mass discs, $\mu\ll1$, the precession frequency is
$\Omega_{\rm pr}=\Omega_\phi-\Omega_R={\mathcal O}(\mu)$ (eq.\
\ref{eq:precrate}) and the surface density $\Sigma_0$ is also ${\mathcal O}(\mu)$. 
For slow modes  $\omega$ is ${\mathcal O}(\mu)$, and since $s=(\omega-m\Omega)/\Omega_R$
is nearly $-m$, the denominator $1-s^2/n^2$ becomes small for $n=m$ and the solution of the 
dispersion relation for a given $m$ is obtained by keeping the dominant $n=m$ term of the 
summation in (\ref{eq:lsk}). Moreover $\sigma_R^2=\frac{1}{2}e_{\rm rms}^2R^2\Omega_R^2$ 
where $e_{\rm rms}$ is the rms eccentricity defined in equation (\ref{eq:erms})---strictly, this is the 
rms eccentricity at a given radius rather than of the whole disc but in our models the rms eccentricity is almost 
independent of radius (Fig.\ \ref{fig:esq}). The dispersion relation for slow modes then simplifies to 
\begin{align}
&\omega=m\Omega_{\rm pr}+\frac{m \pi \Sigma_0|k|}{\Omega_R}{\mathcal
  F}_m \left[(e_{\rm rms}kR)^2/2\right]\notag \\
&\quad \mbox{where}\quad 
{\mathcal  F}_m(\chi)=\frac{2}{\chi}e^{-\chi}I_m(\chi). 
\label{eq:lskslow}
\end{align}
Here all quantities are written in the dimensionless units of
\S\ref{sec:disc-model}. Note that as $z\rightarrow 0$,
$I_m(z)\rightarrow z^m/(2^mm!)$. Thus as the rms eccentricity shrinks
to zero the wavenumber of a slow mode must vary as $k\sim e_{\rm
  rms}^{2(1-m)/(2m-1)}$. In other words, for $m=1$ slow modes have $|kR|\sim 1$
even for cold discs---in this case the use of the WKB approximation
for slow modes is not formally justified, but the results provide a
useful qualitative guide to the behavior of the frequency spectra that
we find using FEM (see further discussion at the end of
\S\ref{sec:prograde-waves}). For $m>1$ slow modes exist but with
wavelengths that shrink as $e_{\rm rms}$ declines. For cold discs
${\mathcal F}_m(\chi)=0$ for $m>1$ so the disc supports only singular
modes at the resonances $\omega=m\Omega_{\rm pr}$. 

In the WKB approximation, disturbances in the disc can be decomposed into wavepackets 
that propagate at the group velocity $\dif \omega/\dif k$ along contours of constant frequency 
$\omega$. These contours are illustrated in Figure \ref{fig:cont} for a disc with mean eccentricity 
$\bar e=0.159$ and $m=1,2$. Wavepackets that propagate along open contours eventually wind up ($|k|\rightarrow\infty$) 
and disappear. Discrete normal modes can arise for closed contours if the appropriate resonance 
condition is satisfied. Consider the case $m=1$ (left panel of Fig.\
\ref{fig:cont}). For the closed contours centered on $|k|=10$, $R=0.6$ the resonance 
condition is (T01)\footnote{The factor of 2 on the left side of equation (56) in T01 is incorrect.}
\begin{equation}
\oint \dif k \, \dif R=2\pi \Big ( n-\frac{1}{2} \Big ),
\label{eq:res}
\end{equation}
where $n=1,2,3,\ldots$ and the integral is taken over the area in $(k,R)$ space enclosed by 
the contour. These modes, called $p$-modes by T01, occur in degenerate pairs, one composed 
of leading and one of trailing waves ($k<0$ and $k>0$ respectively).

Equation (\ref{eq:res}) can be solved numerically to find the
frequencies of the $p$-modes.  These are plotted in the bottom panel
of Figure \ref{fig4} for $m=1$, and are in good qualitative agreement
with the frequencies calculated by FEM. There is similar agreement
between equation (\ref{eq:res}) and FEM for $m=2$ modes.

\end{document}